\documentclass[11pt,onecolumn,a4paper]{article}
\usepackage{amsmath}
\usepackage{amssymb}
\usepackage{graphicx}
\usepackage{bm}
\usepackage{tikz}
\usetikzlibrary{automata, arrows, positioning, calc, bending, decorations.text, decorations.pathreplacing, angles, quotes, fit, patterns}
\usepackage[utf8]{inputenc}
\usepackage{chemformula}
\usepackage{epsfig}
\usepackage{url}
\usepackage{enumitem}
\usepackage{subfig}
\usepackage{epstopdf}
\usepackage{svg}
\usepackage{forest}
\usepackage{algorithm}
\usepackage[noend]{algpseudocode}
\usepackage{booktabs}
\usepackage{pgfplots}
\pgfplotsset{compat=1.18} 
\usepackage{siunitx}
\usepackage{soul}
\usepackage{cleveref}
\usepackage{rotating}
\usetikzlibrary{decorations.pathreplacing,decorations.markings}
\tikzset{
  font=\normalsize,
  red arrow/.style={
    midway,red,sloped,fill, minimum height=1.5cm, single arrow, single arrow head extend=.6cm, single arrow head indent=.25cm,xscale=0.3,yscale=0.15,
    allow upside down
  },
  black arrow/.style 2 args={-stealth, shorten >=#1, shorten <=#2},
  black arrow/.default={1mm}{1mm},
  tree box/.style={draw, rounded corners, inner sep=.3em},
  node box/.style={white, draw=black, text=black, rectangle, rounded corners},
}

\usepackage{xcolor}
\usepackage{mathtools}

\title{\Large{Learning {\it effective  good} variables from physical data}}
\author{Giulio Barletta$^{1}$, Giovanni Trezza$^{1}$, Eliodoro Chiavazzo$^{1}$\thanks{Corresponding author: eliodoro.chiavazzo@polito.it}\\ \small{\emph{$^{1}$Department of Energy, Politecnico di Torino, C.so Duca degli Abruzzi 24, Torino 10129, Italy}}}
\date{}

\usepackage[left=2cm, right=2cm, top=2cm]{geometry}

\begin{document}

\maketitle

\begin{abstract}
We assume that a sufficiently large database is available, where a physical property of interest and a number of associated ruling primitive variables or observables are stored.
We introduce and test two machine learning approaches to discover possible groups or combinations of primitive variables: The first approach is based on regression models whereas the second on classification models.
The variable group (here referred to as the new {\it effective good} variable) can be considered as successfully found, when the physical property of interest is characterized by the following effective invariant behaviour: In the first method, invariance of the group implies invariance of the property up to a given accuracy; in the other method, upon partition of the physical property values into two or more classes, invariance of the group implies invariance of the class.
For the sake of illustration, the two methods are successfully applied to two popular empirical correlations describing the convective heat transfer phenomenon and to the Newton's law of universal gravitation.
\begin{description}
\item[Keywords] \emph{Machine Learning in Physics, Primitive Variable Analysis, Physical Property Invariance, Feature grouping}
\end{description}
\end{abstract}

\section{Introduction\label{intro}}
Theoretical modelling and numerical simulations have become invaluable tools for the current scientific and technological advancement across various fields \cite{rappaz2003numerical}.
In their essence, models make use of a mathematical description of the physical laws by establishing relationships between physical variables.
The latter variables have the key role of describing in a complete and non-redundant manner a system of interest. 
As such, their correct identification is never a trivial task, especially in systems with little prior knowledge \cite{chen2022automated,floryan2022data,eva2023minimal}.

Often, in order to effectively model complex systems, it is favourable to search for a more convenient description with a reduced number of effective variables \cite{chiavazzo2012approximation,chiavazzo2008quasi}.
In other words, the system behaviour is not described by the directly accessible physical variables, but rather by some groups or combinations. 
In this respect, back in the late 19th century, the Buckingham theorem was introduced \cite{rayleigh1892viii, buckingham1914physically,curtis1982dimensional}. 
In the latter approach, based on dimensional analysis, it is possible to mix the primitive physical variables, thus creating fewer dimensionless numbers which are effectively relevant.

Over the years, sophisticated methodologies have been developed also based on machine learning, data mining and other data-driven approaches \cite{chiavazzo2017intrinsic,chiavazzo2014reduced,lin2021data}. 
Some other approaches aim to unlock the discovery of symbolic expressions accurately matching data derived from an unknown function.
This problem has been tackled with a number of methods \cite{mcree2010symbolic, stijven2011separating}, including sparse regression \cite{mcconaghy2011ffx, arnaldo2015building, brunton2016discovering, quade2018sparse}, genetic algorithms \cite{searson2010gptips, dubvcakova2011eureqa, schmidt2009distilling} and physics-inspired algorithms like AI Feynman, introduced by Udrescu \& Tegmark \cite{udrescu2020ai}.
Inspired by the latter, some authors of this work presented a multi-objective optimization procedure for reducing the set of composition-based material descriptors by optimally mixing them in power combination form.
This resulted in improved classification performances, as demonstrated in a case study focused on superconductors \cite{trezza2023leveraging}. 
Moreover, the same procedure was applied by Bonke \emph{et al.} \cite{bonke2023multi} to identify an effective reduced set of variables in a micelle-based photocatalytic system for solar fuels production.
Specifically, the algorithm allowed to analytically mix five physical primitive variables into two synthetic features for the optimal binary separation of the experimental samples according to their performances.
The practical benefit of such approach lies in its capability to replace an experimental sample achieving high performance  with an alternative combination of its elemental components. 
This gives the possibility to reduce the most expensive ones re-balancing the others, at no cost on the overall performance.

In this work we present a general and automated methodology, suitable for regression tasks, able to identify groups and/or group sets of variables in the power form $x_i^{\alpha_1} x_j^{\alpha_2} \cdots x_m^{\alpha_p}$. 
We show its effectiveness on popular thermo-fluid dynamics correlations (i.e. Dittus-Boelter and Gnielinski). 
Furthermore, we also demonstrate that the approach can be easily extended to more general functional form, proving its ability on the Newton's law of universal gravitation.
Also, we employ the optimal feature mixing procedure in ref.~\cite{trezza2023leveraging} for classification tasks, showing its successful application on the former two problems of this study.

The paper is organized in sections as follows. 
%
Section \ref{methods} illustrates the procedure for generating the three datasets employed in this study, based on the three functional forms analyzed here (i.e., Dittus-Boelter correlation, Gnielinski correlation, Newton's law of universal gravitation). 
Furthermore, it presents the methodology for searching good variables in regression models and describes the implementation of classification models for optimal variable mixing towards class separation. 
This section is further subdivided into specific techniques, including the consideration of single and multiple invariant groups in power form, as well as a broader exploration into non-power forms.
%
Section 3 presents numerical examples and discussions related to our study.
We illustrate the application of our methodology on selected cases, including the analysis of the popular Dittus-Boelter and  Gnielinski correlations (each examined from both regression and classification perspectives in power group forms), and Newton's law of universal gravitation (only regression on generic group forms). 
Here, we also provide a comprehensive evaluation of our approach.
%
%
Finally, in Section 4, we draw conclusions providing an overview of the main contributions.



\begin{figure}
    \centering
    \includegraphics[width = \textwidth]{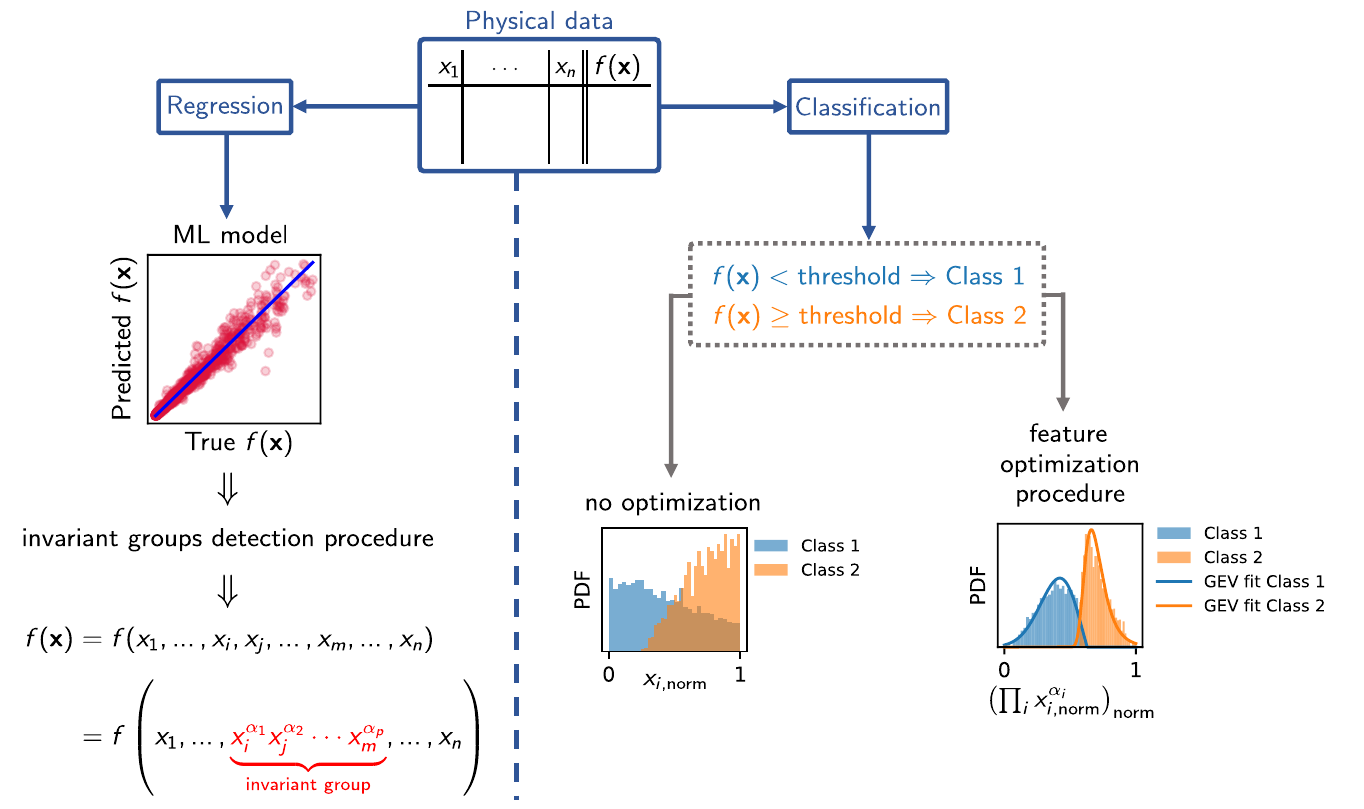}
    \caption{Overview of the protocol used to detect possible symmetries of a target property of interest with respect to its input variables, utilizing only data and ignoring the analytical functional dependence. Two distinct methodologies are presented: the former for identifying, in regression tasks, invariant groups in the form $x_i^{\alpha_1} x_j^{\alpha_2} \cdots x_m^{\alpha_p}$, among others; the latter for identifying, in classification tasks, one or several mixed features as power combinations of the input variables to achieve an optimal class separation. 
    }
    \label{fig:summary}
\end{figure}

\section{Methods} \label{methods}

\subsection{Datasets creation}\label{dataset}

We first create three datasets from physical models to be used to train an test Machine Learning (ML) tools, aiming to predict physical properties of interest.

More specifically, the first two datasets are generated using popular thermo-fluid dynamic correlations (i.e. Dittus-Boelter and Gnielinski correlations) with the target quantity being the Nusselt number.
Those datasets are created starting from the physical properties of 16 real liquids at ambient temperature and pressure~\cite{lide2020crc}, and evaluating the kinematic viscosity $\nu$ and the thermal diffusivity $\kappa$ of each liquid from tabulated data (see Supplementary Note 1).
Each fluid accounts for 500 value combinations of the flow speed $u$, the hydraulic diameter $D$, and the friction factor $f$ (the latter only requested for the Gnielinski correlation).
%
%
The values for these three variables are randomly chosen in the ranges $[0.1,1]$, $[0.01,0.1]$, and $[0.02,0.09]$, respectively. 
This leads to a total of 8,000 samples for each dataset. 
The Nusselt number is thus calculated according to the corresponding equations.
Finally, emulating what is typically experienced in experimental measurements, noise is added on top of the correct target values, sampling 8,000 points $\eta_{i}$ from a Gaussian distribution with mean $\mu=0$ and standard deviation $\sigma=0.1$; the target value of each entry $\mathrm{Nu}_{i}$ is thus multiplied by $(1+\eta_{i})$.

The third dataset is created on the basis of the Newton's law of universal gravitation.
It is constructed in a similar fashion, by generating random values within the range $[10^{16},~10^{18}]$ for the masses $m_{1}$ and $m_{2}$, and within the ranges $[-10^{12},-10^{10}]$ and $[10^{10},10^{12}]$ for the spatial coordinates $x_{1}, y_{1}, z_{1}, x_{2}, y_{2}, z_{2}$. 
Here, the target value is the gravitational force $F_\mathrm{g}$ which is computed with some noise added following the same procedure as above.


\subsection{Searching for good variables by regression models} \label{single}
\subsubsection{Single invariant group in power form}
Let $f(\mathbf{x}) = f(x_{1},\dots,x_{n})$ denote a function depending on $n$ physical variables $x_{1},\dots, x_{n}$. 
If $f(\mathbf{x})$ is invariant with respect to a group of $p$ 
variables in the form $(x_{i}^{\alpha_1}x_{j}^{\alpha_2}\cdots x_{m}^{\alpha_p})$, $\alpha_1, \alpha_2, \dots, \alpha_p \in\mathbb{R}$, when this group is kept at a constant value $\overline{c}$, regardless of the value of each component $x_i, x_j, \dots, x_m$, $f(\mathbf{x})$ does not change. 
Stated differently, the above invariance condition requires: 
\begin{equation}
\label{eq2}
\alpha_1\ln(x_i) + \alpha_2\ln(x_j) + \cdots + \alpha_p\ln(x_m) = c
\end{equation}
($c= \ln(\overline{c})$), which upon differentiation yields:
\begin{equation}
\label{eq3}
    \alpha_1\frac{\mathrm{d}x_{i}}{x_{i}} + \alpha_2\frac{\mathrm{d}x_{j}}{x_{j}} + \cdots + \alpha_p\frac{\mathrm{d}x_{m}}{x_{m}} = 0.
\end{equation}
We can thus construct the $(1 \times p)$ matrix $\mathbf{B}$ at a generic point $\overline{\mathbf{x}_{0}} = (x_{i,0}, x_{j,0}, \dots, x_{m,0})$, as  
\begin{equation}
\label{eq4}
\mathbf{B} = \left[ \frac{\alpha_1}{x_{i,0}}, \frac{\alpha_2}{x_{j,0}}, \cdots, \frac{\alpha_p}{x_{m,0}} \right],
\end{equation}
%
%
%
Let $\mathbf{K}$ be a matrix whose columns form an ortho-normal basis in the null space (or kernel) of $\mathbf{B}$.
Hence, the condition of invariance of $f(\mathbf{x})$ with respect to the group $(x_{i}^{\alpha_1}x_{j}^{\alpha_2}\cdots x_{m}^{\alpha_p})$ can be recast into an orthogonality condition in the configuration space between the normalized gradient of $f(\mathbf{x})$ and each column of \textbf{K}, namely
\begin{equation}
\label{eq5}
    \left({\nabla \Tilde{f}} \right)_{\overline{\mathbf{x}_{0}}}   \cdot  {\mathbf{K}} = 0
\end{equation}
where 
\begin{equation}
\label{eq6}
     {\nabla \Tilde{f}} =\frac{\nabla f}{\mathrm{norm}(\nabla f)}
\end{equation}
Coefficients $\alpha_1, \alpha_2, \dots, \alpha_p$ can be conveniently normalized thus yielding the following algebraic system:
\begin{equation}
\label{eq7}
    \begin{cases}
         \left({\nabla \Tilde{f}} \right)_{\overline{\mathbf{x}_{0}}}   \cdot  {\mathbf{K}} = 0
        \\
        \alpha_1^{2}+\alpha_2^{2}+ \cdots +\alpha_p^{2}=1
    \end{cases}
\end{equation}
%
%
%
If the non-linear system in Eq.~(\ref{eq7}) is satisfied for the same exponents ($\overline{\alpha_1}$, $\overline{\alpha_2}$, \dots, $\overline{\alpha_p}$) over all the domains of the features ($x_{i}$, $x_{j}$, \dots, $x_{m}$), the group $x_{i}^{\overline{\alpha_1}}x_{j}^{\overline{\alpha_2}}\cdots x_{m}^{\overline{\alpha_p}}$ does represent an intrinsic variable and $f(\mathbf{x})$ is invariant with respect to that group.

\subsubsection{Multiple concurrent invariant groups in power form} \label{couples}

Clearly, the above procedure can be extended to a function $f(\mathbf{x})$ being invariant with respect to a higher number of feature groups, with each group even sharing some of the primitive physical variables.
Without loosing generality, and for the sake of simplicity, we limit this generalized description to sets consisting of two concurrent groups.

Let $f(\mathbf{x}) = f(x_{1},\dots,x_{n}) = f(x_{1},\dots,x_{i}^{\alpha_1}x_{j}^{\alpha_2}\cdots x_{m}^{\alpha_p}, x_{j}^{\beta_1}x_{k}^{\beta_2}\cdots x_{r}^{\beta_q}, \dots, x_{n})$ 
denote a function where two groups share a generic primitive variable $x_j$. 
In this case, the above procedure applied to individual groups $x_{i}^{\alpha_1}x_{j}^{\alpha_2} \cdots x_{m}^{\alpha_p}$ and $x_{j}^{\beta_1}x_{k}^{\beta_2} \cdots x_{r}^{\beta_q}$ independently is not suitable any longer and requires a generalization as discussed below.

To investigate the invariance with respect to both groups, we now construct the $(2\times l)$ 
matrix $\mathbf{B}$:
\begin{equation}
\label{eq8}
\mathbf{B} =
\begin{bmatrix}
    \frac{\alpha_1}{x_{i,0}} & \frac{\alpha_2}{x_{j,0}} & 0 & \cdots & \frac{\alpha_p}{x_{m,0}} & 0 \\
    0 & \frac{\beta_1}{x_{j,0}} & \frac{\beta_2}{x_{k,0}} & \cdots & 0 & \frac{\beta_q}{x_{r,0}}
\end{bmatrix},
\end{equation}
%
with $ \max(p, q) \le l \le p+q$, $\mathbf{K}$ being a matrix whose columns represent an ortho-normal basis of the null space of $\mathbf{B}$.
%

As done above, the condition of invariance requires that the normalized gradient of $f(\mathbf{x})$ is orthogonal to each column of the kernel matrix \textbf{K} (see also Eq.~(\ref{eq5})).

Adding the normalization condition of the coefficients $\alpha_1, \alpha_2, \dots, \alpha_p$ and $\beta_1, \beta_2, \dots, \beta_q$, we obtain:
\begin{equation}
\label{eq10}
   \begin{cases}
\left({\nabla \Tilde{f}} \right)_{\overline{\mathbf{x}_{0}}}   \cdot  {\mathbf{K}} = 0
        \\
        \alpha_1^{2}+\alpha_2^{2}+\dots+\alpha_p^{2}=1
        \\
        \beta_1^{2}+\beta_2^{2}+\dots+\beta_q^{2}=1
    \end{cases}
\end{equation}

with the partial derivatives being evaluated in $\overline{\mathbf{x}_{0}}$.
If the non-linear system in Eq.~(\ref{eq10}) is satisfied for the same exponents ($\overline{\alpha_1}$, \dots, $\overline{\alpha_p}$, $\overline{\beta_1} \dots, \overline{\beta_q}$) in the entire feature domain ($x_{i}$, $x_{j}$, \dots, $x_{r}$), the two groups $x_{i}^{\overline{\alpha_1}}x_{j}^{\overline{\alpha_2}} \cdots x_{m}^{\overline{\alpha_p}}$ and $x_{j}^{\overline{\beta_1}}x_{k}^{\overline{\beta_2}} \cdots x_{r}^{\overline{\beta_q}}$ are intrinsic variables as the function $f(\mathbf{x})$ is invariant with respect to both groups.

It is worth stressing that more general cases with three or more concurrent invariant groups will imply additional rows for the matrix $\mathbf{B}$.
Furthermore, in general, the non-linear system (\ref{eq10}) is not necessarily closed (see subsection \ref{dnn}).


\subsubsection{Further generalization to non power forms} \label{general}
The invariant group/set identification procedure, introduced for groups in the power form, can be generalized to other functional relationships.
For the sake of illustration, in the following we restrict to a single invariant group.

Let $f(\mathbf{x}) = f(x_{1},\dots,x_{n})$ denote a function invariant with respect to a group involving $p$ variables according to a generic functional dependence $g(x_i, x_j, \dots, x_m)$.
When such group is a constant $\overline{c}$, even varying its components $x_i, x_j, \dots, x_m$ separately, $f(\mathbf{x})$ does not change. 
This yields:
\begin{equation}
\label{eq11}
    g(\underbrace{x_{i}, x_{j}, \dots, x_{m}}_{p}) = \overline{c}
\end{equation}
which translates into:
\begin{equation}
\label{eq12}
    \frac{\partial g}{\partial x_{i}} \mathrm{d}x_{i} + \frac{\partial g}{\partial x_{j}} \mathrm{d}x_{j} + \dots + \frac{\partial g}{\partial x_{m}} \mathrm{d}x_{m} = 0.
\end{equation}
We can thus construct the $(1\times p)$ matrix $\mathbf{B}$ at a generic point $\overline{\mathbf{x}_{0}} = (x_{i,0}, x_{j,0}, \dots, x_{m,0})$, as
\begin{equation}\label{eq13}
\mathbf{B} = \left[ \frac{\partial g}{\partial x_{i}}, \frac{\partial g}{\partial x_{j}}, \dots, \frac{\partial g}{\partial x_{m}} \right]
\end{equation}
Let $\mathbf{K}$ be a matrix whose colomns represent an orthonormal basis of the null space of $\mathbf{B}$.
Applying the invariance condition of Eq.~(\ref{eq5}) leads to a non-linear system analogous to Eq.~(\ref{eq7}).
Notably, the procedure illustrated here can be further extended to sets of groups in any functional form, following a reasoning similar to that applied in the above subsection \ref{couples}.

\subsubsection{Regression model and procedure implementation} \label{dnn}
In this study, we assume that a database of physical data is available.
Each sample in our database consist of $n$ features $(x_1, \dots, x_n)$ corresponding to a target quantity $f(\mathbf{x})$.
We further assume that the target quantity possibly depends on some invariant groups expressed for instance in the power form.
We thus try to detect such invariance following the above procedure.

As already described, this requires the evaluation of the gradient $\nabla f(\mathbf{x})$. 
To this end, $f(\mathbf{x})$ can be conveniently approximated with a Deep Neural Network (DNN), allowing the computation of that gradient by means of automatic differentiation.
All the DNNs of this study are trained and validated over the 85\% of the corresponding databases – of which the 85\% is used for the training and the remaining 15\% for the validation – and tested over the remaining 15\%. 
The DNN structure, as reported in Supplementary Note 2, is used for all the examples of this study.
Upon network training and validation, automatic differentiation is utilized to calculate the gradient of the function at a specific point $\mathbf{x}_{0}$ within its domain. 
Here, the variables in the examined group are randomly chosen within their domains, while the values of the remaining variables are held constant at their averages in the original database.

The function's gradient is computed and applied in Eq. (\ref{eq7}) and Eq. (\ref{eq10}), depending on the specific case. 
Subsequently, the system is solved using a least-squares optimization method that utilizes the Trust Region Reflective~\cite{branch1999subspace} algorithm, which is suitable also for under-determined (non closed) systems, that may arise during the process of identifying sets of groups.

This strategy is repeated with 20 different values of $\mathbf{x}_{0}$ and iterated multiple times, each time updating the initial guess to the mean value of the exponents $(\overline{\alpha_1}, \overline{\alpha_2}, \dots, \overline{\alpha_p})$ averaged over the 20 solutions found in the previous iteration.
%
%
%

\begin{figure}
    \centering
    \includegraphics[width=.85\textwidth]{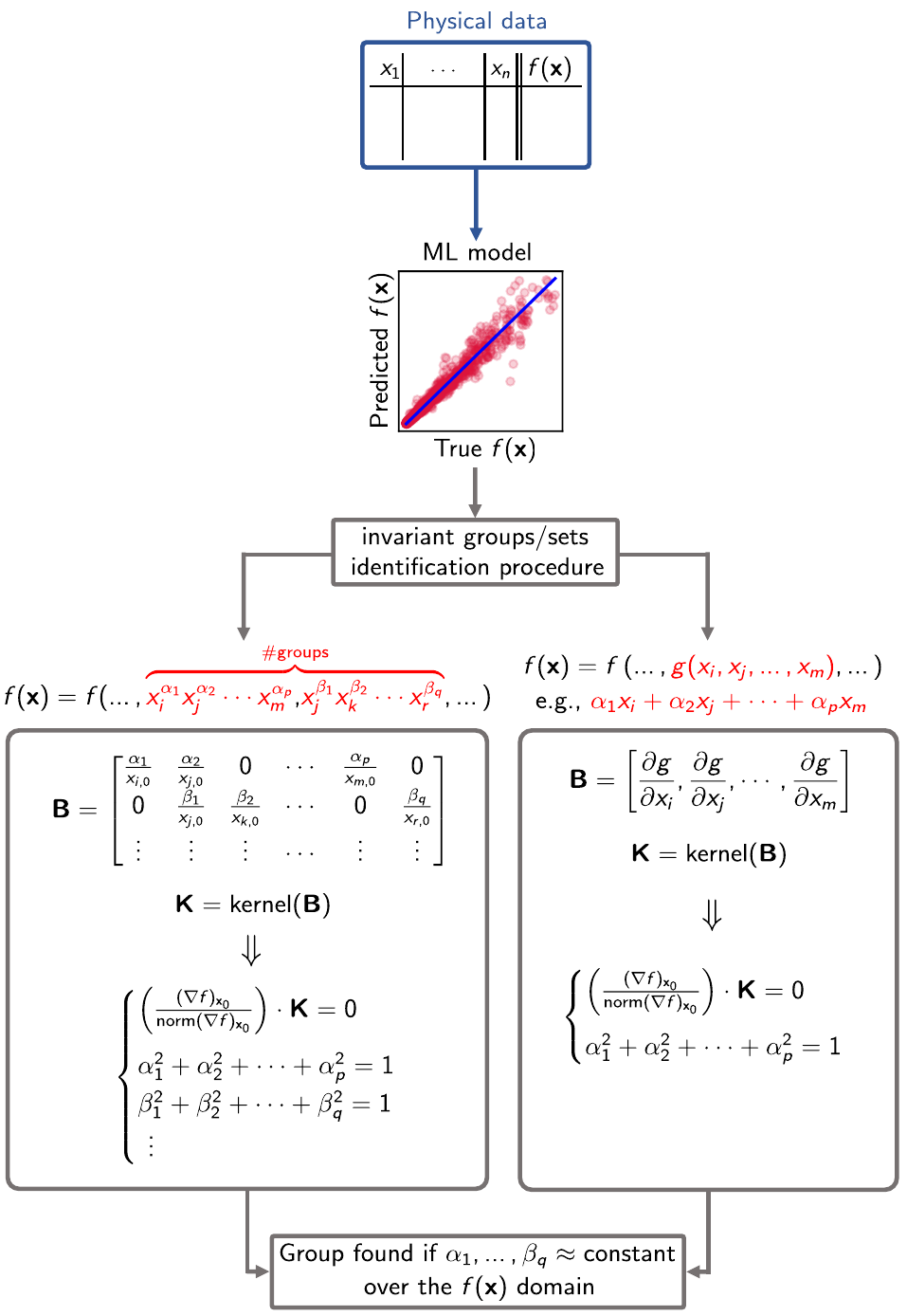}
    \caption{Overview of the procedure for identifying invariant groups/sets. A regression model is trained on the physical data and used to compute the gradient of the objective function in a point $\mathbf{x}_{0}$. The matrix \textbf{B} is constructed according to the functional structure of the investigated group/set, and its kernel \textbf{K} is computed. Finally, the condition of invariance between gradient and kernel is coupled to the normalization conditions of the coefficients. If the resulting non-linear system is satisfied for the same coefficients over the $f(\mathbf{x})$ domain, the group/set is an intrinsic variable and $f(\mathbf{x})$ is invariant with respect to it. 
    }
    \label{fig:summary_regression}
\end{figure}

An overview of the methodology for identifying invariant groups/sets is depicted in Fig.~\ref{fig:summary_regression}.

\subsection{Searching for good variables by classification models} \label{classification}
%

The methodology introduced above, based on regression models, is able to identify groups and/or group sets of physical variables upon which a generic function $f(\mathbf{x})$ depends.
If successful, this approach efficiently reduces the number of such primitive variables by optimizing their combination. 

However, a similar goal can be achieved by means of an entirely different approach based on classification models, and it has been initially proposed by Trezza \& Chiavazzo in \cite{trezza2023leveraging}.
The latter methodology, briefly reviewed in the following, enables the optimal mixing of primitive features by performing classification and subsequent optimal class separation of the available data samples.

Let $x_{1},\dots,x_{n}$ denote $n$ features. 
Let $(\Tilde{x}_{1},\dots,\Tilde{x}_{n})$ be the corresponding dimensionless quantities
\begin{equation}
\label{eq15}
    \Tilde{x}_{i} = \frac{x_{i}-x_{i,\mathrm{min}}}{x_{i,\mathrm{max}}-x_{i,\mathrm{min}}}+1
\end{equation}
normalized by construction within the range $[1, 2]$ to avoid singularities in the procedure below, where $x_{i,\mathrm{min}}$ and $x_{i,\mathrm{max}}$ represent the minimum and the maximum observed values for the $i$th feature over the training set, respectively.
It is possible to synthetically create a set of $m\ll n$ mixed features $(y_{1},\dots,y_{m})$ as
\begin{equation}
\label{eq16}
    y_{j}=\prod_{i=1}^{n} \Tilde{x}_{i}^{\alpha_{ij}}
\end{equation}
with $\{\alpha_{ij}\}$ being a $(n \times m)$ matrix estimated by means of a multi-objective optimization algorithm, as described below. 
Finally, the new variables $y_{j}$ can be normalized within the interval $[0, 1]$ according to
\begin{equation}
\label{eq17}
    \Tilde{y}_{j} = \frac{y_{j}-y_{j,\mathrm{min}}}{y_{j,\mathrm{max}}-y_{j,\mathrm{min}}}.
\end{equation}

The main idea is that the matrix {$\alpha_{ij}$} shall be selected following an optimization criterion attempting the largest possible separation between two (or more) different classes partitioning the values of a physical property of interest. 
Clearly, this can be accomplished by maximizing a certain distance between the classes.
However, a multi-objective optimization procedure could also be pursued.
For instance in a binary classification, the matrix {$\alpha_{ij}$} in Eq.~(\ref{eq16}) can be chosen to lie on the Pareto front while simultaneously pursuing: (i) maximization of a carefully selected distance between the two classes; (ii) minimization of the norm of the covariance matrix of the first class distribution; (iii) minimization of a norm of the covariance matrix of the second class distribution.

The main rationale behind the minimization of a norm of the covariance matrix for class distribution (along with a distance between classes) is the aim of possibly obtaining smooth distribution functions that can be analytically best fitted.
Following the latter idea, below we will pursue multi-objective optimization for all the examined cases and provide some optimal solutions from Pareto fronts.


In this study, we adopt genetic algorithms for optimization whereas the Bhattacharyya distance between the histograms of the two equally binned classes \cite{bhattacharyya1943measure,bhattacharyya1946measure} is evaluated to be maximized during the multi-objective optimization. 
However, as shown in ref.~\cite{trezza2023leveraging}, other options for class distance may be considered, such as the Wasserstein distance \cite{villani2009optimal} or the averaged number within a fixed radius of nearest neighbors of one class to each samples of the other class.
%
Furthermore, herein we utilize variable power-form mixingoutlined in Eq.~(\ref{eq16}). Nonetheless, alternative grouping options (e.g., linear mixing as employed in ref.~\cite{trezza2023leveraging}) are also possible with the overall methodology remaining unchanged.
Also, the procedure can be further generalized to a number of classes $>2$. 

\begin{figure}
    \centering
    \includegraphics[width = 0.7\textwidth]{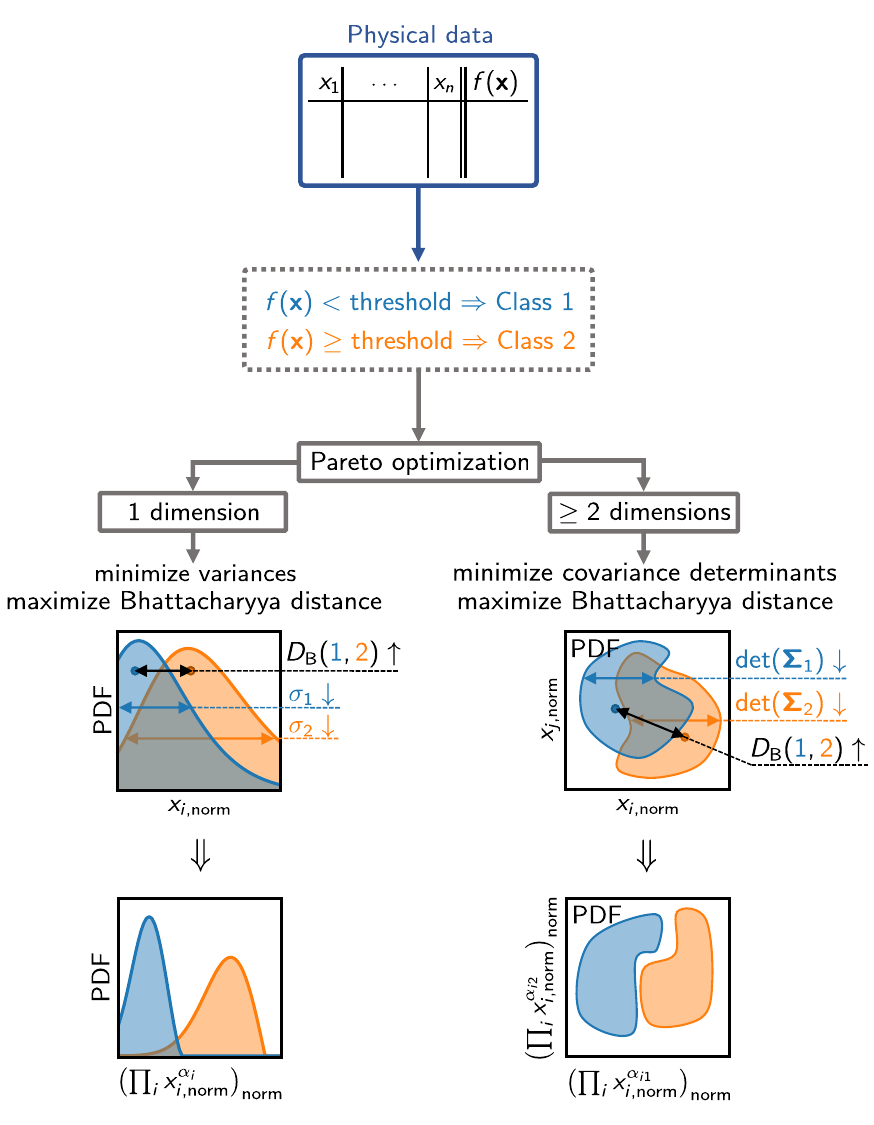}
    \caption{Overview of the procedure to identify optimal mixed variables for class separation. Threshold values are chosen to divide the physical data in classes. A Pareto optimization is performed to construct a reduced set of synthetic features that simultaneously maximizes the Bhattacharyya distance between the classes and (i) minimizes the variances of the class distributions, in the one dimensional case, or (ii) minimizes the determinants of the covariance matrix of the class distributions, in the multi dimensional case.}
    \label{fig:classification}
\end{figure}

In this case, the genetic algorithm aims at simultaneously maximizing the pairwise distances between the classes \cite{trezza2023leveraging}.
%
%
For the remaining two objectives, in one-dimensional cases, a numerical estimate of the standard deviation of the binned data in the two classes is computed. 
In two-dimensional (or higher) cases, the determinant of the covariance matrix can be utilized.
%
From the practical standpoint, considering a dataset of physical data samples characterized by features $x_1, \dots, x_n$ and their corresponding response $f(\mathbf{x})$, such dataset is partitioned into two classes based on a carefully chosen threshold for $f(\mathbf{x})$. 
After the aforementioned pre-processing steps, a genetic algorithm optimization is employed to identify the Pareto front. 
%
%
This optimization concurrently seeks to minimize the variances of the two classes (in one dimension) or the determinants of the covariances of the two classes (in two or more dimensions), while also maximizing the Bhattacharyya distance between the classes. 
A summary of this procedure is illustrated in Fig.~\ref{fig:classification}
%
%
%
%

\section{Numerical examples and discussion} \label{examples}


\subsection{Dittus-Boelter equation} \label{dittus}

The Dittus-Boelter correlation for a fluid undergoing heating is expressed by the following equation:

\begin{equation}
\label{eq18}
    \mathrm{Nu} = 0.023\mathrm{Re}_{D}^{4/5}\mathrm{Pr}^{0.4}
\end{equation}
where $\mathrm{Re}_{D} = \frac{ud}{\nu}$ represents the Reynolds number and $\mathrm{Pr} = \frac{\nu}{\kappa}$ denotes the Prandtl number. 
The equation can be rewritten by substituting the dimensionless quantities $Re$ and $Pr$, thus obtaining
\begin{equation}
\label{eq19}
    \mathrm{Nu} = 0.023\left(\frac{ud}{\nu}\right)^{4/5}\left(\frac{\nu}{\kappa}\right)^{0.4} = 0.023\frac{u^{0.8}d^{0.8}}{\nu^{0.4}\kappa^{0.4}}.
\end{equation}
It is easy to verify that the objective value is invariant with respect to all the possible combinations of input features, $u, d, \nu, \kappa$, either couples or triplets.


\subsubsection{Use of regression models} \label{dittus_reg}

\begin{figure}
    \centering
    \includegraphics[width = \textwidth]{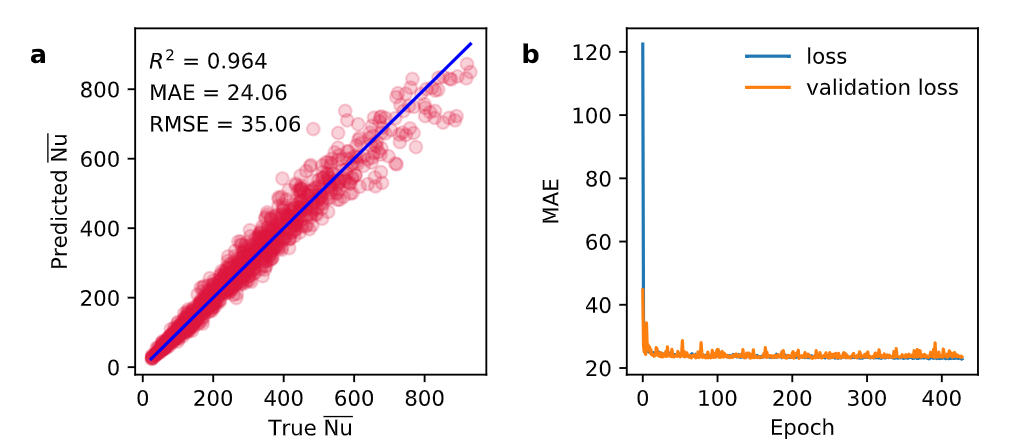}
    \caption{Results of the DNN regression model for the noised Nusselt number $\overline{\mathrm{Nu}}$ in the Dittus-Boelter correlation. \textbf{a} Predictions over the testing set, and \textbf{b} corresponding loss curves for the DNN model. Model performances are shown in terms of coefficient of determination $R^{2}$, mean absolute error (MAE), and root mean squared error (RMSE).}
    \label{fig:dittus_model}
\end{figure}

We attempt to recover the symmetries of the Nusselt number with respect to binary and ternary groups in the form $x_{i}^{\alpha_1}x_{j}^{\alpha_2}$ and $x_{i}^{\alpha_1}x_{j}^{\alpha_2}x_{k}^{\alpha_3}$ only relying on noised data and by adopting the methodology illustrated in subsection \ref{single}.
As outlined above, this requires the evaluation of the gradient of the noised Nusselt number with respect to the input features, namely $\nabla \overline{\mathrm{Nu}}(u,d,\nu,\kappa)$, where for each sample $i$, $\overline{\mathrm{Nu}}_i = \mathrm{Nu}_i(1+\eta_i)$ (see Methods for details). This can be conveniently computed by automatic differentiation over a DNN model approximating $\overline{\mathrm{Nu}}(u,d,\nu,\kappa)$.
As input features of the DNN, the four variables $u,d,\nu,\kappa$ are used. 
The dataset is thus divided into three parts: (i) a training set, (ii) a validation set used to detect potential overfitting, and (iii) a testing set for the comprehensive evaluation of model performance. 
Fig.~\ref{fig:dittus_model}a showcases the model predictions over the testing set, while in Fig.~\ref{fig:dittus_model}b the corresponding loss across epochs is depicted. 

Notably, the model is highly predictive, with coefficient of determination $R^2=0.964$ over the testing set, with no evidence of overfitting observed.
The model is thus fed to the optimization algorithm. 
%
For each of the ten expected invariant groups (six binary and four ternary), the algorithm estimates the normalized exponents $\alpha_1, \alpha_2, \alpha_3$ 20 times. 
Table \ref{table3} shows such true normalized exponents $\alpha_1, \alpha_2, \alpha_3$, together with the means $\mu_{\alpha_1}, \mu_{\alpha_2}, \mu_{\alpha_3}$ and the standard deviations $\sigma_{\alpha_1}, \sigma_{\alpha_2}, \sigma_{\alpha_3}$ over those 20 evaluations.
%
%
%
Our findings indicate that the method correctly identifies invariant groups in both couple and triplet forms. 
Furthermore, it demonstrates the ability to estimate normalized exponents $\alpha_1, \alpha_2, \alpha_3$ for individual primitive variables within each group, with relative percent errors not exceeding 6.0\% for couples and 16.7\% for triplets.
Clearly, we notice that normalized exponents are obtained up to the sign: the exponents determined for the pair $(u, d)$ are negative instead of the expected positive values according to the Dittus-Boelter equation.
%

In Table \ref{table3}, we label as \emph{found} all the groups with standard deviations $\sigma_{\alpha_1}, \sigma_{\alpha_2}, \sigma_{\alpha_3} \leq 0.2$, since low values of $\sigma$ imply that the results for the 20 evaluations are consistent.
We also label as \emph{reliable} all the groups where none of the evaluated exponents approach 0 or 1, since such cases would correspond to trivial solutions.
Remarkably, all six binary groups and all four ternary groups of input features are found to comply with the condition of invariance, and the results for all of them are reliable.
\begin{table}
    \centering
    \caption{Normalized exponents $\alpha_1, \alpha_2, \alpha_3$ for the Dittus-Boelter correlation together with their average estimates over 20 evaluations $\mu_{\alpha_1}, \mu_{\alpha_2}, \mu_{\alpha_3}$ and the corresponding standard deviations $\sigma_{\alpha_1}, \sigma_{\alpha_2}, \sigma_{\alpha_3}$. Found groups refer to low standard deviation, reliable groups refer to average far from 1 and 0.}
    \label{table3}
    \begin{tabular}{crrrrrrrrrcc}
    \toprule
         \textbf{Group} & $\alpha_1$ & $\alpha_2$ & $\alpha_3$ & ${\mu_{\alpha_1}}$ & ${\sigma_{\alpha_1}}$ &  ${\mu_{\alpha_2}}$ & ${\sigma_{\alpha_2}}$ &   ${\mu_{\alpha_3}}$ & ${\sigma_{\alpha_3}}$ & \textbf{Found} & \textbf{Reliable}  \\
         \midrule
         $(u,d)$ & 0.71 & 0.71 & - & -0.67 & 0.12 & -0.73 & 0.10 & - & - & yes & yes\\
         $(u,\nu)$ & 0.89 & -0.45 & - & 0.90 & 0.03 & -0.44 & 0.07 & - & -  & yes & yes\\
         $(u,\kappa)$ & 0.89 & -0.45 & - & 0.90 & 0.02 & -0.42 & 0.04 & - & - & yes & yes \\
         $(d,\nu)$ & 0.89 & -0.45 & - & 0.88 & 0.06 & -0.46 & 0.06 & - & - & yes & yes \\
         $(d,\kappa)$ & 0.89 & -0.45 & - & 0.88 & 0.04 & -0.47 & 0.07 & - & - & yes & yes \\
         $(\nu,\kappa)$ & -0.71 & -0.71 & - & -0.70 & 0.09 & -0.71 & 0.09 & - & - & yes & yes \\         
         $(u,d,\nu)$ & 0.67 & 0.67 & -0.33 & 0.67 & 0.04 & 0.65 & 0.05 & -0.34 & 0.09 & yes & yes \\ 
         $(u,d,\kappa)$ & 0.67 & 0.67 & -0.33 & 0.66 & 0.03 & 0.66 & 0.05 & -0.34 & 0.07 & yes & yes \\ 
         $(u,\nu,\kappa)$ & 0.82 & -0.41 & -0.41 & 0.88 & 0.05 & -0.34 & 0.09 & -0.35 & 0.07 & yes & yes \\ 
         $(d,\nu,\kappa)$ & 0.82 & -0.41 & -0.41 & 0.86 & 0.03 & -0.37 & 0.10  & -0.34 & 0.07 & yes & yes \\ 
         \bottomrule
    \end{tabular}
\end{table}

\subsubsection{Use of classification models} \label{dittus_class}

\begin{figure}
    \centering
    \includegraphics[width = \textwidth]{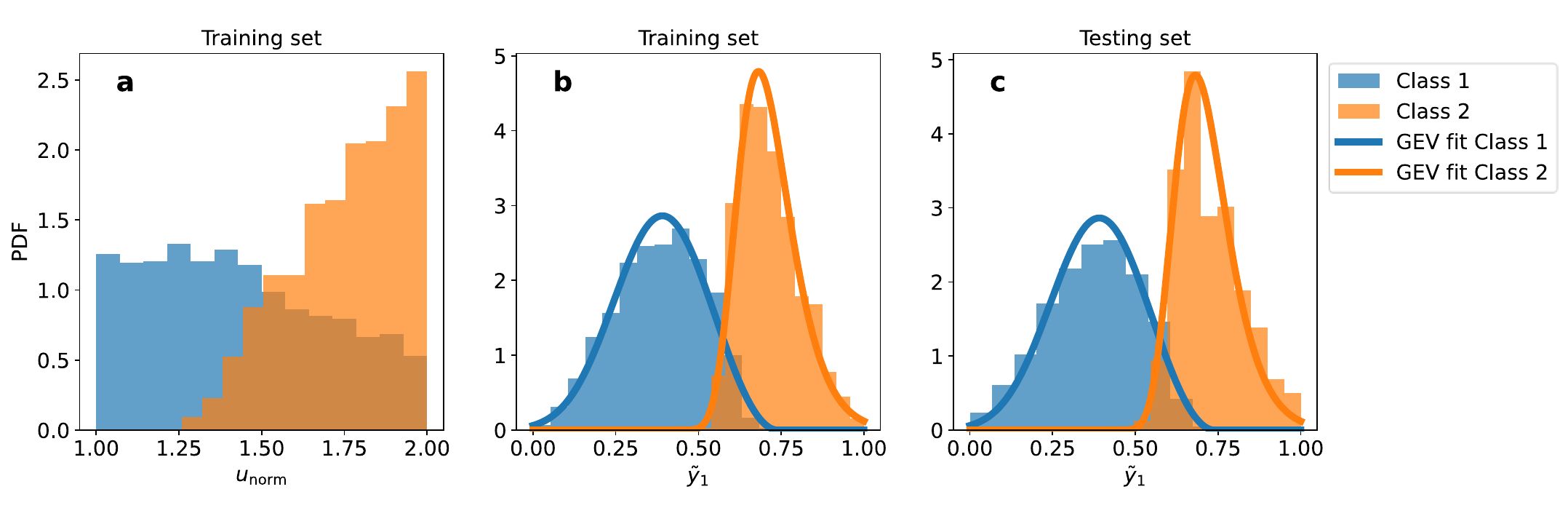}
    \caption{One dimensional example for classification on the Dittus-Boelter correlation. \textbf{a} PDFs over binned data of the training set for the two classes ($\overline{\mathrm{Nu}}<395$ and $\overline{\mathrm{Nu}} \geq 395$) reported against the normalized flow velocity. \textbf{b} PDFs over binned data of the training set for the two classes reported against the mixed feature $y_{1}$, constructed according to Eq.~(\ref{eq16}) and choosing the point of the Pareto front with the least overlapping of the two classes according to the Bhattacharyya distance, along with a GEV analytical fitting of the two binnings. \textbf{c} PDFs over binned data of the testing set for the two classes reported against the same mixed feature $y_{1}$ together with the same GEV fittings of the \textbf{b} subfigure. The mixed variable $\Tilde{y}_1$ shown here is referred exclusively to this Dittus-Boelter one dimensional optimization.}
    \label{fig:dittus_class}
\end{figure}

In a second attempt of reducing the number of input variables, we investigate the possible existence of symmetries for classification, aiming at the construction of $m$ mixed features in the form $y_{j}=\prod_{i=1}^{n} \Tilde{x}_{i}^{\alpha_{ij}}$, where $(\Tilde{x}_{1},\dots,\Tilde{x}_{n})$ are the primitive variables properly normalized within the interval [1, 2] and $\{\alpha_{ij}\}$ denotes an $n \times m$ matrix optimally estimated, as detailed in subsection \ref{classification}. 
Such features are finally properly normalized as $\Tilde{y}_j$ in the interval [0, 1].
In the case of the Dittus-Bolter, we set class 1 for samples with $\overline{\mathrm{Nu}}<395$ and class 2 for samples with $\overline{\mathrm{Nu}}\geq395$.
We thus create a single mixed feature ($m=1$) according to the above procedure.
Fig.~\ref{fig:dittus_class}a reports the PDF binning of the training set data over the two classes, against the normalized flow velocity $u_{\textrm{norm}}$, while Fig.~\ref{fig:dittus_class}b shows the same PDFs against the normalized mixed feature $\Tilde{y}_1$.
%
%
It is noteworthy to observe that when represented against the primitive feature, the two classes exhibit considerable overlap, whereas the same two classes appear well separated when plotted against the mixed feature.
As a result, it is particularly convenient to attempt an analytical best-fitting of the two distributions depicted in Fig.~\ref{fig:dittus_class}b approximated by a Generalized Extreme Value (GEV) distribution, with density
\begin{equation}
\label{eq21}
    g(\Tilde{y}_{1}) = \frac{1}{\sigma} \left( 1+\omega \frac{\Tilde{y}_{1}-\mu}{\sigma} \right)^{-\frac{\omega+1}{\omega}}  \exp{\left( -\left( 1+\omega \frac{\Tilde{y}_{1}-\mu}{\sigma} \right)^{-1/\omega} \right)}.
\end{equation}
The fitting is performed by means of the SciPy Python package~\cite{2020SciPy-NMeth}. 
The specific computed GEV distribution for samples with $\overline{\mathrm{Nu}}<395$ has factors $\mu = 0.338$, $\sigma = 0.137$, and $\omega = 0.342$, whereas the GEV distribution for samples with $\overline{\mathrm{Nu}} \geq 395$ has factors $\mu = 0.675$, $\sigma = 0.077$, and $\omega = 0.074$.
Fig.~\ref{fig:dittus_class}c shows the PDFs over the binned data of the testing set reported against the same mixed feature $\Tilde{y}_{1}$, along with the GEV fittings computed on the training set. 
Notably, the classes are still well separated, with a good agreement between the GEV distributions and the testing set densities.

\begin{figure}
    \centering
    \includegraphics[width = 0.6\textwidth]{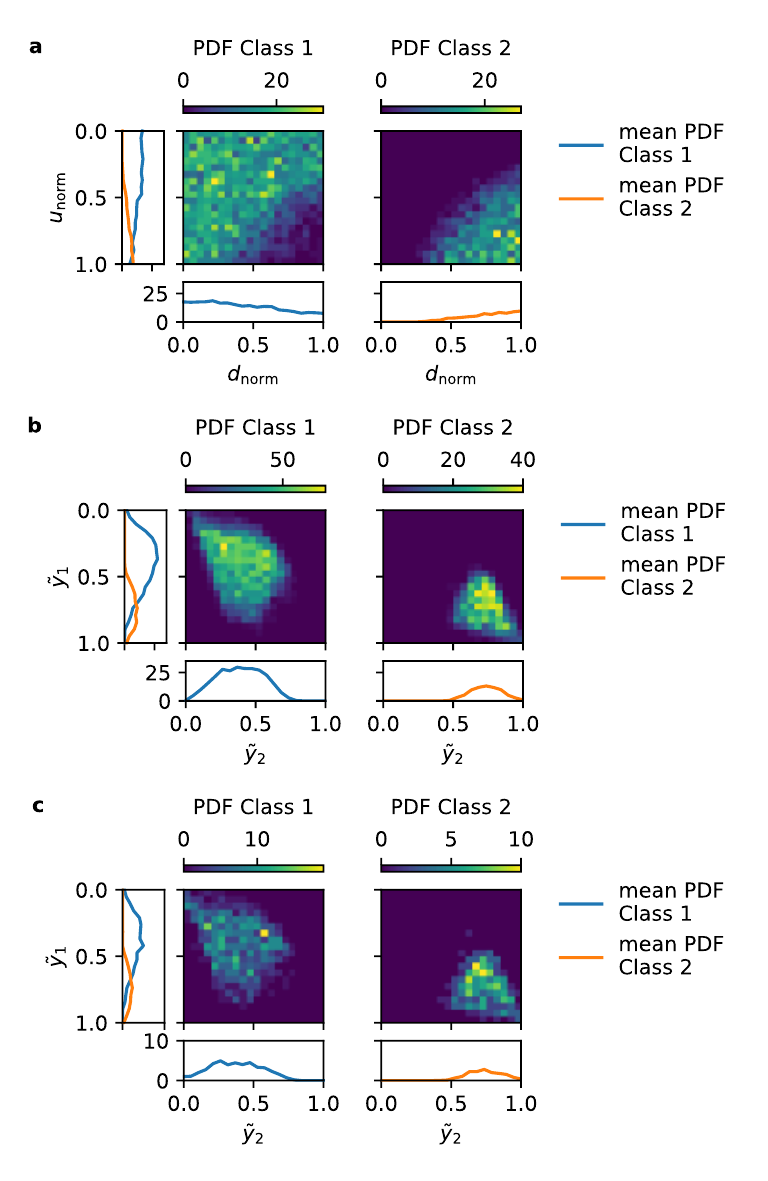}
    \caption{Two dimensional example for the classification on the Dittus-Boelter correlation. \textbf{a} PDFs over binned data of the training set for the two classes ($\overline{\mathrm{Nu}}<395$ and $\overline{\mathrm{Nu}} \geq 395$) reported against the normalized flow velocity $u_{\mathrm{norm}}$ and the normalized hydraulic diameter $d_{\mathrm{norm}}$. \textbf{b} PDFs over binned data of the training set for the two classes reported against the mixed features $\Tilde{y}_{1}, \Tilde{y}_{2}$, constructed according to Eq.~(\ref{eq16}) and choosing the point of the Pareto front with the least overlapping of the two classes according to the Bhattacharyya distance. \textbf{c} PDFs over binned data of the testing set for the two classes reported against the same mixed features $\Tilde{y}_{1}, \Tilde{y}_{2}$. The mixed variables $\Tilde{y}_1, \Tilde{y}_2$ shown here are referred exclusively to this Dittus-Boelter two dimensional optimization.}
    \label{fig:dittus_2features}
\end{figure}

Furthermore, we make an attempt at the creation of two mixed normalized features $\Tilde{y}_1, \Tilde{y}_2$ ($m=2$) with the same classes. 
Specifically, Fig.~\ref{fig:dittus_2features}a shows the PDF two dimensional binning of the training set data over the two classes, against the normalized flow velocity $u_\mathrm{norm}$ and  the hydraulic diameter $d_\mathrm{norm}$.
Fig.~\ref{fig:dittus_2features}b shows the same PDFs against the normalized mixed features $\Tilde{y}_{1}, \Tilde{y}_{2}$ constructed according to Eq.~(\ref{eq16}) by power combination of the four primitive variables and choosing the point of the Pareto front with the least overlapping of the two distributions. 
Similarly to the one dimensional case, the classes appear well separated when plotted against the mixed features, whereas there is a wide overlap when plotted against the primitive variables.
Fig.~\ref{fig:dittus_2features}c shows the PDFs over the binned data of the testing set reported against the same mixed features. 
The two classes appear still well separated.

\begin{figure}
    \centering
    \includegraphics[width = \textwidth]{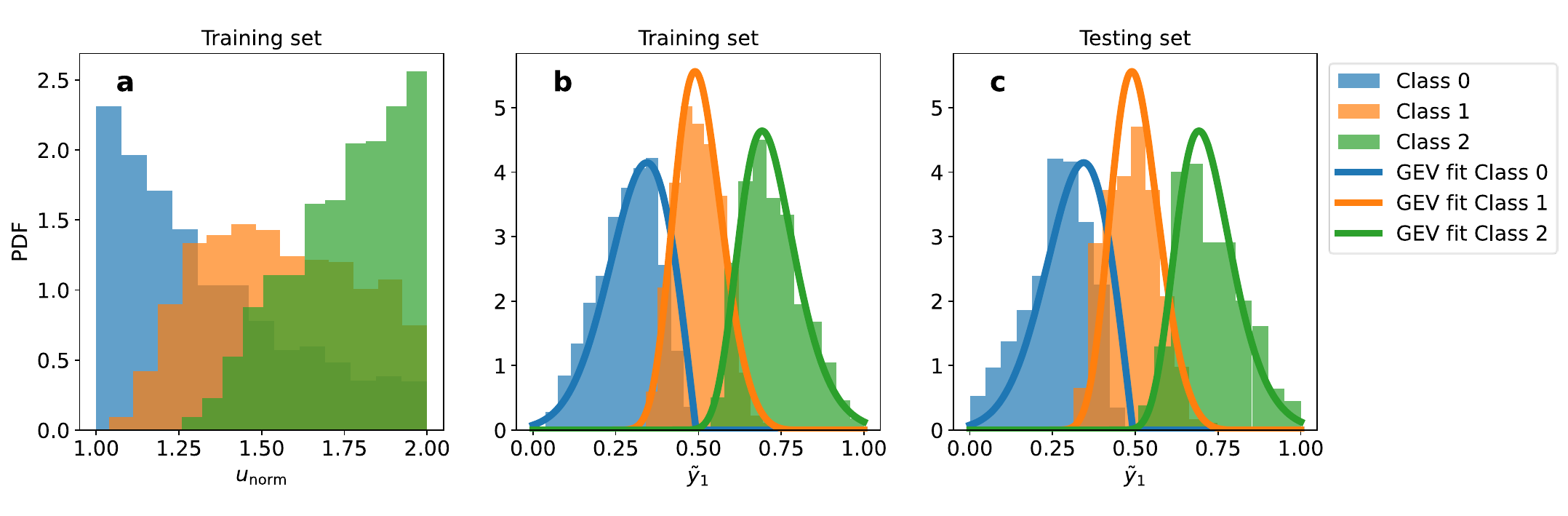}
    \caption{One dimensional example for the ternary classification on the Dittus-Boelter. \textbf{a} PDFs over binned data of the training set for the three classes ($\overline{\mathrm{Nu}}<197.5$, $197.5 \leq \overline{\mathrm{Nu}} < 395$, and  $\overline{\mathrm{Nu}}\geq 395$) reported against the normalized flow velocity $u_{\mathrm{norm}}$. \textbf{b} PDFs over binned data of the training set for the three classes reported against the mixed feature $\Tilde{y}_{1}$, constructed according to Eq.~(\ref{eq16}) and choosing the point of the Pareto front with the least overlapping of the three classes according to the Bhattacharyya distance, along with a GEV analytical fitting of the three binnings. \textbf{c} PDFs over binned data of the testing set for the three classes reported against the same mixed feature $\Tilde{y}_{1}$ together with the same GEV fittings of the \textbf{b} subfigure. The mixed variable $\Tilde{y}_1$ shown here is referred exclusively to this Dittus-Boelter one dimensional optimization.}
    \label{fig:dittus_3class}
\end{figure}

Finally, we aim at the construction of $m=1$ mixed feature for separating the data samples into three classes: class 0 for $\overline{\mathrm{Nu}}<197.5$, class 1 for $197.5\leq\overline{\mathrm{Nu}}<395$, class 2 for $\overline{\mathrm{Nu}}\geq395$. The multi-objective optimization is performed aiming at concurrently maximizing the pairwise distances between the classes. 
Fig.~\ref{fig:dittus_3class} shows the separation in classes reported against the normalized flow velocity $u_\mathrm{norm}$ and the new mixed feature $\Tilde{y}_{1}$: in line with previous cases, the substantial overlap observed when representing classes against the primitive feature is significantly reduced when using the optimized mixed feature.
The bet-fitting GEV distributions for the three classes are computed according to Eq.~(\ref{eq21}) over the training set samples. 
Specifically the GEV associated with Class 0 has factors $\mu = 0.284$, $\sigma = 0.103$, $\omega = 0.494$, the GEV for Class 1 has factors $\mu = 0.475$, $\sigma = 0.067$, $\omega = 0.193$, and the GEV for Class 2 has factors $\mu = 0.684$, $\sigma = 0.080$, $\omega = 0.103$. 
When plotting the PDFs over the binned data of the testing set against the mixed feature, the classes appear well separated, and there is good agreement between the GEV distributions and the testing set densities.

\subsection{Gnielinski correlation} \label{gnielinski}

The Gnielinski correlation for turbulent flow in tubes is expressed by the equation
\begin{equation}
\label{eq22}
    \mathrm{Nu} = \frac{(f/8)(\mathrm{Re}_{D}-1000)\mathrm{Pr}}{1+12.7(f/8)^{1/2}(\mathrm{Pr}^{2/3}-1)}
\end{equation}
where $\mathrm{Re}_{D} = \frac{ud}{\nu}$ represents the Reynolds number, $\mathrm{Pr} = \frac{\nu}{\kappa}$ is the Prandtl number and $f$ denotes the friction factor. 
The equation can be reformulated by substituting the dimensionless quantities $Re$ and $Pr$, resulting in:
\begin{equation}
\label{eq23}
    \mathrm{Nu} = \frac{(f/8) \left(\frac{ud}{\nu}-1000 \right)\left(\frac{\nu}{\kappa}\right)}{1+12.7(f/8)^{1/2}\left(\left(\frac{\nu}{\kappa}\right)^{2/3}-1\right)}.
\end{equation}
The Nusselt number is invariant only with respect to the combination of flow velocity and hydraulic diameter $(u,d)$ with real exponents $(1,1)$, normalized exponents $(0.707,0.707)$. 

\subsubsection{Use of regression models} \label{gniel_reg}

\begin{figure}
    \centering
    \includegraphics[width = \textwidth]{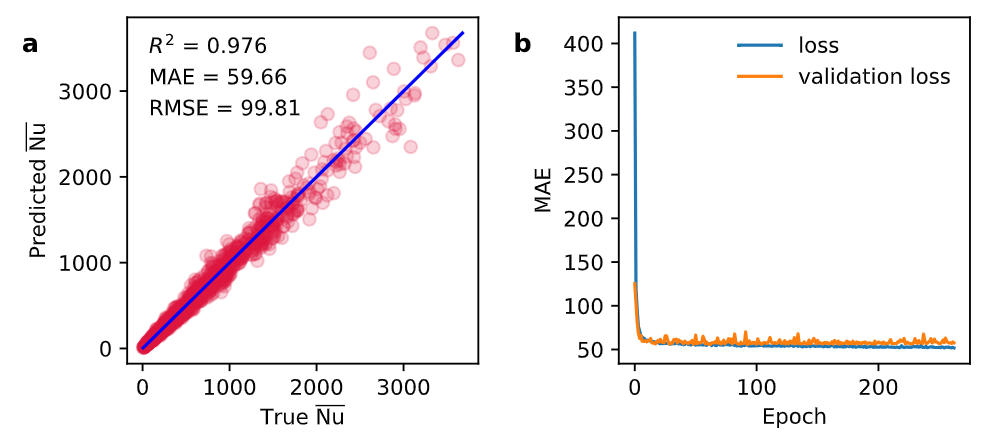}
    \caption{Results of the DNN regression model for the noised Nusselt number $\overline{\mathrm{Nu}}$ in the Gnielinski correlation. \textbf{a} Predictions over the testing set, and \textbf{b} corresponding loss curves for the DNN model. Model performances are shown in terms of coefficient of determination $R^{2}$, mean absolute error (MAE), and root mean squared error (RMSE).}
    \label{fig:gniel_model}
\end{figure}

Here, we first attempt the detection of possible symmetries of the noised Nusselt number $\overline{\mathrm{Nu}}$ adopting the methodology proposed in subsection \ref{couples}.
We get access to the gradient $\nabla \overline{\mathrm{Nu}}(u,d,\nu,\kappa,f)$ by means of automatic differentiation over a DNN approximating $\overline{\mathrm{Nu}}(u,d,\nu,\kappa,f)$.
Fig.~\ref{fig:gniel_model}a shows the model predictions over the testing set, while in Fig.~\ref{fig:gniel_model}b the corresponding loss across epochs is depicted. 
The model is highly predictive, with coefficient of determination $R^2=0.976$ over the testing set, and no evidence of overfitting is observed.
The trained model is thus fed to the optimization algorithm, looking for binary groups in the form $x_i^{\alpha_1} x_j^{\alpha_2}$.
The average $\mu_{\alpha_1}, \mu_{\alpha_2}$ estimates of the exponents $\alpha_1, \alpha_2$, together with their standard deviations  $\sigma_{\alpha_1}, \sigma_{\alpha_2}$ are reported in Table \ref{table4}. 

Specifically, the procedure correctly identifies the expected group $(u,d)$, whereas for most other groups the results correspond to either the trivial solution (i.e., the absolute value of one of the exponents is close to 1 and the other is close to 0), or at least one of the evaluated values presents too high variance (i.e., $\sigma_{\alpha_i} > 0.2$), as for $(\kappa,f)$.
%
%
Remarkably, for the remaining cases (namely $(u,\nu)$, $(u,f)$, $(d,\nu)$, $(d,f)$, $(\nu,\kappa)$, $(\nu,f)$) the procedure identifies groups complying with the invariance condition, albeit not explicitly expected in the Gnielinski correlation.
%
%
We thus concluded that the suggested procedure has detected a \emph{local} invariance and this can be numerically verified as follows (here we limit to groups of two variables only).  
%
%
First, the features not appearing in the group are kept constant.
Second, a random sample $i$ in the dataset is considered for the evaluation of the quantity $\Tilde{c}=x_{1,i}^{\alpha_1} x_{2,i}^{\alpha_2}$.
Third, a vector $\overline{\mathbf{x}_{1}}$ of evenly spaced points in the domain of the first feature is created. 
As such, the array corresponding to the second feature is obtained as $\overline{\mathbf{x}_2} = (\Tilde{c}\overline{\mathbf{x}_1}^{-\alpha_1})^{1/\alpha_2}$. 
A new dataset is thus constructed, with the variables out of the group being constant and with the variables in the group being replaced by $\overline{\mathbf{x}_1}$, $\overline{\mathbf{x}_2}$. 
For all those samples, the response value $f(\mathbf{x})$ is computed. 
The local invariance is demonstrated when $f(\mathbf{x})$ is approximately constant over all the new constructed dataset. 
%
%
Further details can be found in Supplementary Note 3. 
%
%
\begin{sidewaystable}
\fontsize{8}{9}\selectfont
    \centering
    \caption{Normalized exponents $\alpha_1, \alpha_2, \alpha_3, \beta_1, \beta_2$ for the Gnielinski correlation, together with their average estimates over 20 evaluations $\mu_{\alpha_1}$, $\mu_{\alpha_2}$, $\mu_{\alpha_3}$, $\mu_{\beta_1}$, $\mu_{\beta_2}$ and the corresponding standard deviations $\sigma_{\alpha_1}$, $\sigma_{\alpha_2}$, $\sigma_{\alpha_3}$, $\sigma_{\beta_1}$, $\sigma_{\beta_2}$. Found groups/sets refer to low standard deviation, reliable groups refer to average far from 1 and 0.}
    \begin{tabular}{crrrrrrrrrrrrrrrcc}
         \toprule
         \textbf{Group/Set} & $\alpha_1$ & $\alpha_2$ & $\alpha_3$ & $\beta_1$ & $\beta_2$ & ${\mu_{\alpha_1}}$ & ${\sigma_{\alpha_1}}$ & ${\mu_{\alpha_2}}$ & ${\sigma_{\alpha_2}}$ & ${\mu_{\alpha_3}}$ & ${\sigma_{\alpha_3}}$ & ${\mu_{\beta_1}}$ & ${\sigma_{\beta_1}}$ & ${\mu_{\beta_2}}$ & ${\sigma_{\beta_2}}$ & \textbf{Found} & \textbf{Reliable} \\
         \midrule
         $(u,d)$ & 0.707 & 0.707 & - & - & - & -0.727 & 0.050 & -0.682 & 0.055 & - & - & - & - & - & - & yes & yes\\
         $(u,\nu)$ & - & - & - & - & - & 0.830 & 0.041 & -0.553 & 0.062 & - & - & - & - & - & - & yes & yes\\
         $(u,\kappa)$ & - & - & - & - & - & 0.944 & 0.025 & -0.322 & 0.072 & - & - & - & - & - & - & yes & no\\
         $(u,f)$ & - & - & - & - & - & -0.829 & 0.082 & -0.521 & 0.132 & - & - & - & - & - & - & yes & yes\\
         $(d,\nu)$ & - & - & - & - & - & 0.838 & 0.045 & -0.538 & 0.073 & - & - & - & - & - & - & yes & yes\\
         $(d,\kappa)$ & - & - & - & - & - & -0.928 & 0.053 & 0.350 & 0.117 & - & - & - & - & - & - & yes & no\\
         $(d,f)$ & - & - & - & - & - & 0.862 & 0.080 & 0.457 & 0.104 & - & - & - & - & - & - & yes & yes \\
         $(\nu,\kappa)$ & - & - & - & - & - & -0.896 & 0.065 & -0.405 & 0.170 & - & - & - & - & - & - & yes & yes \\
         $(\nu,f)$ & - & - & - & - & - & 0.720 & 0.070 & -0.687 & 0.075 & - & - & - & - & - & - & yes & yes \\
         $(\kappa,f)$ & - & - & - & - & - & 0.518 & 0.142 & -0.839 & 0.285 & - & - & - & - & - & - & no & - \\
         \midrule
         $[(u,\nu),(\nu,\kappa)]$ & 0.707 & -0.707 & - & 0.707 & -0.707 & 0.709 & 0.066 & -0.698 & 0.067 & - & - & -0.692 & 0.017 & 0.722 & 0.016 & yes & yes\\
         $[(d,\nu),(\nu,\kappa)]$ & 0.707 & -0.707 & - & 0.707 & -0.707 & 0.702 & 0.073 & -0.704 & 0.076 & - & - & -0.700 & 0.019 & 0.714 & 0.019 & yes & yes\\
         $[(u,d),(u,\nu)]$ & - & - & - & - & - & 0.260 & 0.042 & -0.965 & 0.010 & - & - & -0.903 & 0.055 & 0.314 & 0.287 & yes & no\\
         $[(u,d),(d,f)]$ & - & - & - & - & - & -0.541 & 0.059 & -0.838 & 0.042 & - & - & -0.569 & 0.020 & 0.822 & 0.014 & yes & yes\\
         $[(u,\nu),(\nu,f)]$ & - & - & - & - & - & 0.858 & 0.024 & -0.512 & 0.044 & - & - & -0.183 & 0.047 & 0.982 & 0.006 & yes & no\\
         $[(u,f),(f,d)]$ & - & - & - & - & - & -0.632 & 0.080 & -0.768 & 0.069 & - & - & -0.495 & 0.058 & 0.867 & 0.035 & yes & yes\\
         $[(d,f),(f,\nu)]$ & - & - & - & - & - & 0.849 & 0.018 & -0.527 & 0.028 & - & - & -0.883 & 0.038 & 0.463 & 0.064 & yes & yes\\        
         \midrule
         $[(u,d,\nu),(\nu,\kappa)]$ & 0.577 & 0.577 & -0.577 & 0.707 & -0.707 & 0.587 & 0.045 & 0.562 & 0.073 & -0.572 & 0.078 & 0.700 & 0.023 & -0.713 & 0.022 & yes & yes\\
         $[(u,d,f),(f,\nu)]$ & - & - & - & - & - & 0.573 & 0.005 & 0.544 & 0.031 & -0.612 & 0.032 & 0.833 & 0.023 & -0.551 & 0.033 & yes & yes\\
         $[(u,\kappa,f),(\kappa, \nu)]$ & - & - & - & - & - & 0.422 & 0.055 & 0.721 & 0.034 & -0.540 & 0.077 & 0.606 & 0.029 & -0.794 & 0.021 & yes & yes\\
         \bottomrule
    \end{tabular}
    \label{table4}
\end{sidewaystable}


Furthermore, armed with the same DNN regression model, we try to detect invariance of the Nusselt number with respect to sets groups in the form $x_{i}^{\alpha_1}x_{j}^{\alpha_2}x_{k}^{\alpha_3}$ and $x_{k}^{\beta_1}x_{l}^{\beta_2}$ employing the procedure presented in subsection \ref{couples}.
%
\Cref{table4} reports also the results for selected sets of two feature couples, including sets appearing explicitly in the Gnielinski correlation -- i.e., $[(u,\nu),(\nu,\kappa)]$ and $[(d,\nu),(\nu,\kappa)]$ -- and for selected sets of features comprising a triplet and a couple -- e.g., $[(u,d,\nu),(\nu,\kappa)]$, which is extremely relevant as it comprises the Reynolds and the Prandtl numbers.
For all the mentioned sets the procedure correctly identifies the exponents that make the groups compliant with the condition of invariance. 
Indeed, the normalized exponents obtained with the procedure for the sets $[(u,\nu),(\nu,\kappa)]$ and $[(d,\nu),(\nu,\kappa)]$ are $[(0.709, -0.698), (-0.692, 0.722)]$ and $[(0.702, -0.704), (-0.700, 0.714)]$ respectively, whereas the analytical solution is $[(0.707,-0.707),(0.707,-0.707)]$ for both the cases. 
The normalized exponents obtained for the set $[(u,d,\nu),(\nu,\kappa)]$ are $[(0.587, 0.562, -0.572), (0.700, -0.713)]$, whereas the analytical solution is $[(0.577,0.577,-0.577),(0.707,-0.707)]$.
%
%

\Cref{table4} also reports some identified sets that locally comply with the condition of invariance (see Supplementary Note 3 for more details), together with some randomly selected sets for which the procedure does succeed at finding the invariance property, but for which the results are close to the trivial solution, meaning that at least one of the evaluated exponents approaches either 0 or $\pm1$, thus being not reliable.
%
%
Following the same approach above, Table \ref{table4} labels as \emph{found} groups with low standard deviations and as \emph{reliable} all the groups in which none of the evaluated exponents approaches 0 or $\pm1$. 
Remarkably, the analytically evident couple $(u, d)$, sets of two couples $[(u,\nu),(\nu,\kappa)]$ and $[(d,\nu),(\nu,\kappa)]$, and set comprising a triplet and a couple $[(u,d,\nu),(\nu,\kappa)]$, are identified. 
Furthermore, eight additional couples are found that locally comply with the condition of invariance, six of them being reliable, and only one couple is not found. 
In the case of the selected sets of two feature couples, five more sets of two couples locally comply with the condition of invariance, only two of them not being reliable. 
Finally, the two additional sets comprising a triplet and a couple are reliably found to represent a local invariance.
%

\subsubsection{Use of classification models} \label{gniel_class}

\begin{figure}
    \centering
    \includegraphics[width = \textwidth]{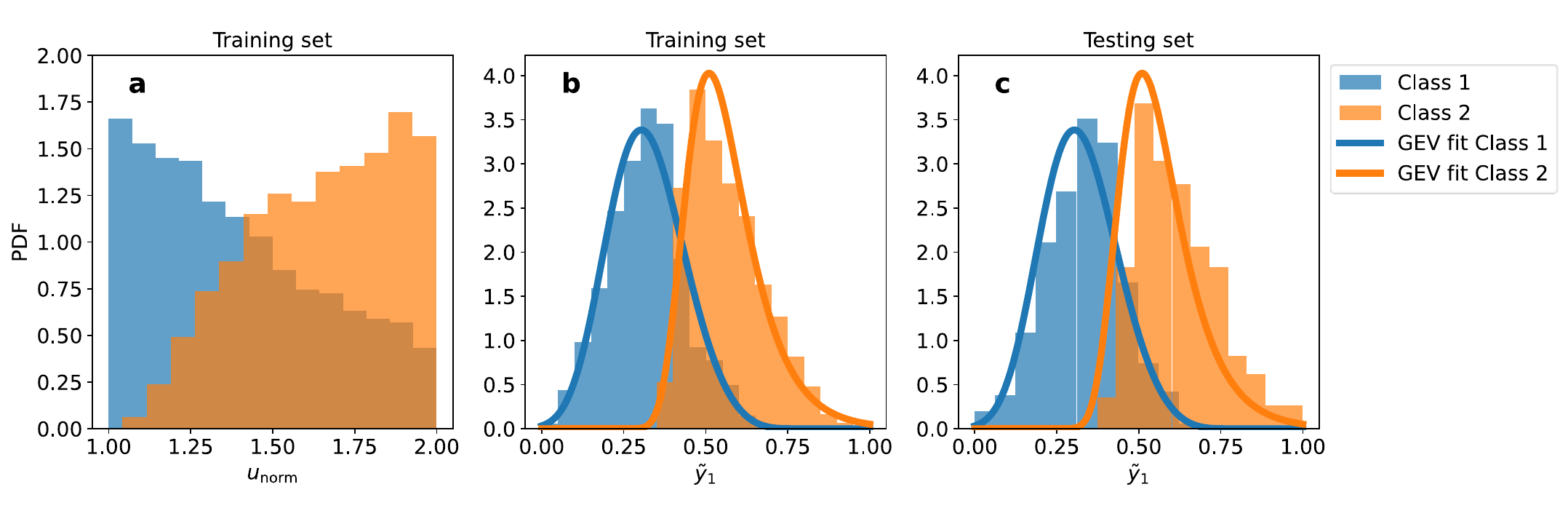}
    \caption{One dimensional example: \textbf{a} PDFs over binned data of the training set for the two classes ($\overline{\mathrm{Nu}}<500$ and $\overline{\mathrm{Nu}}\geq 500$) reported against the normalized flow velocity. \textbf{b} PDFs over binned data of the training set for the two classes reported against the mixed feature $y_{1}$, constructed according to Eq.~(\ref{eq16}) and choosing the point of the Pareto front with the least overlapping of the two classes according to the Bhattacharyya distance, along with a GEV analytical fitting of the two binnings. \textbf{c} PDFs over binned data of the testing set for the two classes reported against the same mixed feature $y_{1}$ together with the same GEV fittings of the \textbf{b} subfigure. The mixed variable $\Tilde{y}_1$ shown here is referred exclusively to this Gnielinski one dimensional optimization.}
    \label{fig:gniel_class}
\end{figure}

\begin{figure}
    \centering
    \includegraphics[width = 0.6\textwidth]{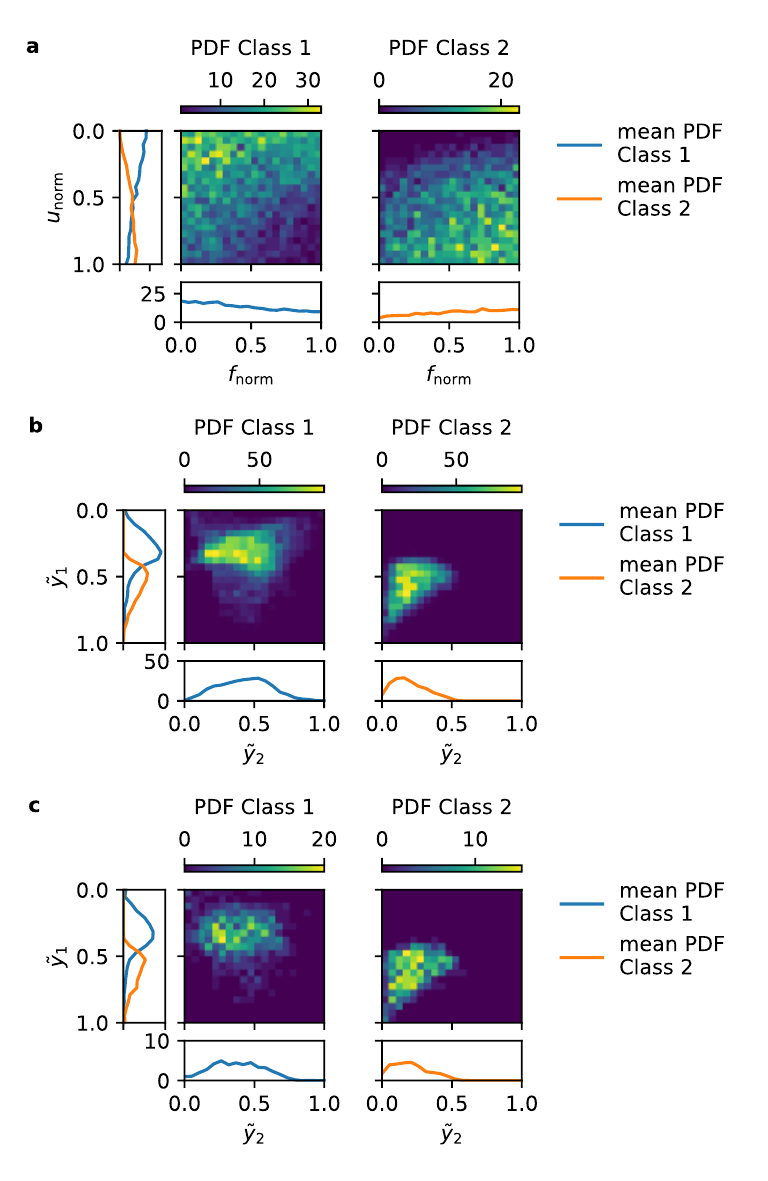}
    \caption{Two dimensional example for the classification on the Gnielinski correlation: \textbf{a} PDFs over binned data of the training set for the two classes ($\overline{\mathrm{Nu}}<500$ and $\overline{\mathrm{Nu}}\geq 500$) reported against the normalized flow velocity $u_{\mathrm{norm}}$ and friction factor $f$. \textbf{b} PDFs over binned data of the training set for the two classes reported against the mixed features $\Tilde{y}_{1}, \Tilde{y}_{2}$, constructed according to Eq.~(\ref{eq16}) and choosing the point of the Pareto front with the least overlapping of the two classes according to the Bhattacharyya distance. \textbf{c} PDFs over binned data of the testing set for the two classes reported against the same mixed features $\Tilde{y}_{1}, \Tilde{y}_{2}$. The mixed variables $\Tilde{y}_1, \Tilde{y}_2$ shown here are referred exclusively to this Gnielinski two dimensional optimization.}
    \label{fig:gniel_2features}
\end{figure}

\begin{figure}
    \centering
    \includegraphics[width = \textwidth]{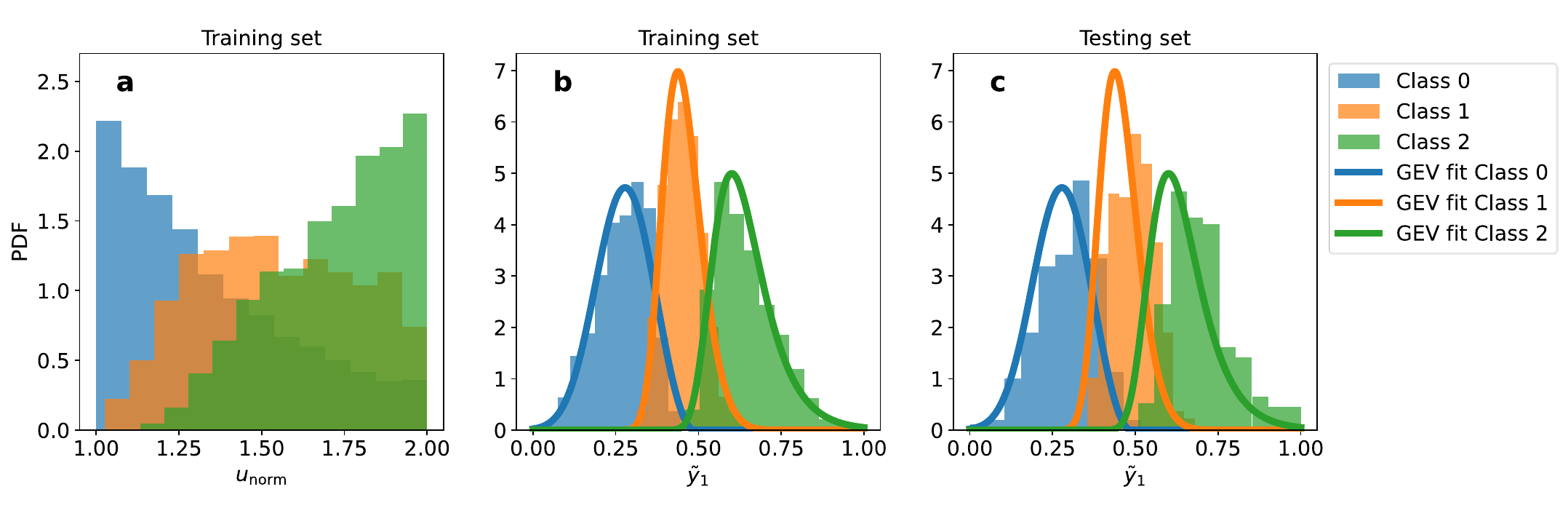}
    \caption{One dimensional example for the ternary classification on the Gnielinski correlation. \textbf{a} PDFs over binned data of the training set for the three classes ($\overline{\mathrm{Nu}}<400$, $400 \leq \overline{\mathrm{Nu}} < 900$, and  $\overline{\mathrm{Nu}}\geq 900$) reported against the normalized flow velocity $u_{\mathrm{norm}}$. \textbf{b} PDFs over binned data of the training set for the three classes reported against the mixed feature $\Tilde{y}_{1}$, constructed according to Eq.~(\ref{eq16}) and choosing the point of the Pareto front with the least overlapping of the three classes according to the Bhattacharyya distance, along with a GEV analytical fitting of the three binnings. \textbf{c} PDFs over binned data of the testing set for the three classes reported against the same mixed feature $\Tilde{y}_{1}$ together with the same GEV fittings of the \textbf{b} subfigure. The mixed variable $\Tilde{y}_1$ shown here is referred exclusively to this Gnielinski one dimensional optimization.}
    \label{fig:gniel_3class}
\end{figure}

In a second attempt at reducing the number of input variables, we aim at the construction of mixed optimized features as power combinations of the normalized primitive variables for classification, according to Eq.~(\ref{eq16}).
%
%
In the case of Gnielinski correlation, we set class 1 for samples with $\overline{\mathrm{Nu}}<500$ and class 2 for samples with $\overline{\mathrm{Nu}}\geq500$.
%
Once the optimization routine finds the Pareto front, the mixed features are created using the point of the Pareto front with the least overlapping of the two classes according to the Battacharyya distance.
Fig.~\ref{fig:gniel_class}a reports the PDF binning of the training set data over the two classes, against the normalized flow velocity $u_{\textrm{norm}}$, while Fig.~\ref{fig:gniel_class}b shows the same PDFs against the normalized mixed feature $\Tilde{y}_1$; still, when represented against the primitive variable, the two classes exhibit significant overlap, whereas they appear distinctly separated when plotted against the mixed feature. 
Two bet-fitting GEV distributions are computed for the two classes according to Eq.~(\ref{eq21}) for samples in class 1 with factors $\mu = 0.276$, $\sigma = 0.112$, $\omega = 0.223$; for samples in class 2 with factors $\mu = 0.510$, $\sigma = 0.091$, $\omega = 0.000$. Fig.~\ref{fig:gniel_class}c shows the PDFs over the binned data of the testing set reported against the same mixed feature $\Tilde{y}_{1}$, along with the GEV fittings computed on the training set. Notably, the classes are still well separated, with a good agreement between the GEV distributions and the testing set densities. 
Furthermore, we attempt to create two mixed normalized features $\Tilde{y}_1, \Tilde{y}_2$ with the same classes. Fig.~\ref{fig:gniel_2features}a shows the PDF two dimensional binning of the training set data over the two classes, against the normalized flow velocity $u_\mathrm{norm}$ and friction factor $f_\mathrm{norm}$. Fig.~\ref{fig:gniel_2features}b shows the same PDFs against the mixed features $\Tilde{y}_{1}, \Tilde{y}_{2}$ constructed according to Eq.~(\ref{eq16}) by power combination of the five relevant features and choosing the point of the Pareto front with the least overlapping of the classes. Similarly to the one dimensional case, the classes appear again well separated when plotted against the mixed features, whereas there is a wide overlap when plotted against the primitive variables. \Cref{fig:gniel_2features}c shows the PDFs over the binned data of the testing set, reported against the same mixed features $\Tilde{y}_{1}, \Tilde{y}_{2}$; the two classes still appear well separated.
Finally, we aim at the construction of $m=1$ mixed feature for separating the data samples into three classes: class 0 for $\overline{\mathrm{Nu}}<400$, class 1 for $400\leq\overline{\mathrm{Nu}}<900$, class 2 for $\overline{\mathrm{Nu}}\geq 900$. The multi-objective optimization is performed aiming at concurrently maximizing the pairwise distances between the classes. 
Fig.~\ref{fig:gniel_3class} shows the separation in classes reported against the normalized flow velocity $u_\mathrm{norm}$ and the new mixed feature $\Tilde{y}_{1}$: in line with previous cases, the considerable overlap observed when representing classes against the primitive feature is significantly reduced when using the optimized mixed feature.
The bet-fitting GEV distributions for the three classes are computed according to Eq.~(\ref{eq21}) over the training set samples; specifically the GEV associated with Class 0 has factors $\mu = 0.245$, $\sigma = 0.083$, $\omega = 0.346$, the GEV for Class 1 has factors $\mu = 0.430$, $\sigma = 0.053$, $\omega = 0.143$, and the GEV for Class 2 has factors $\mu = 0.599$, $\sigma = 0.074$, $\omega = 0.017$. When plotting the PDFs over the binned data of the testing set against the mixed feature, the classes appear well separated, and there is good agreement between the GEV distributions and the testing set densities.

\subsection{Newton's law of universal gravitation} \label{newton}

The module of the interacting force $F_\mathrm{g}$ for two objects of masses $m_1, m_2$ is expressed via Newton's law of universal gravitation
\begin{equation}
\label{eq}
    F_{\mathrm{g}} = \frac{Gm_{1}m_{2}}{(x_{1}-x_{2})^{2}+(y_{1}-y_{2})^{2}+(z_{1}-z_{2})^{2}},
\end{equation}
where $G$ is the gravitational constant, $(x_1,y_1,z_1)$ and $(x_2,y_2,z_2)$ are the coordinates of the centers of their masses. 

\begin{figure}
    \centering
    \includegraphics[width = \textwidth]{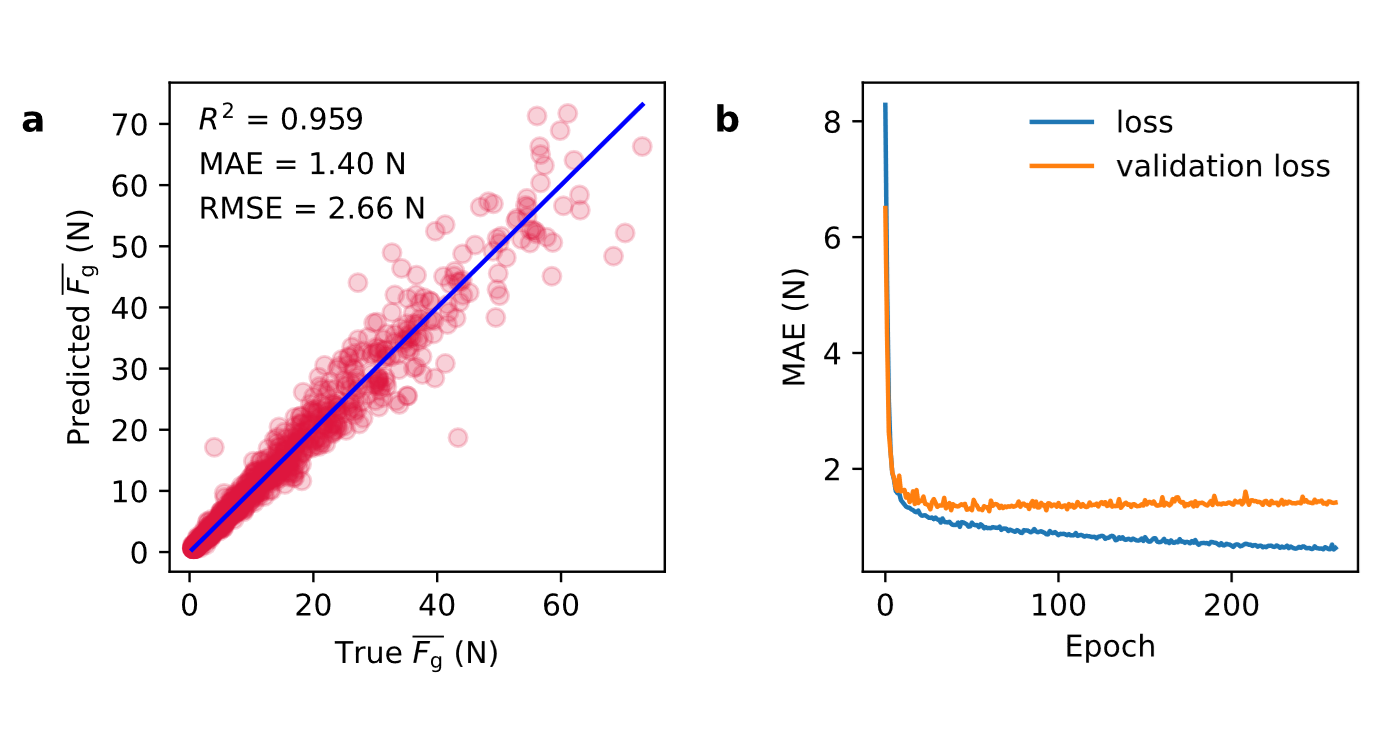}
    \caption{Results of the DNN regression model for the noised gravitational force $F_\mathrm{g}$. \textbf{a} Predictions over the testing set, and \textbf{b} corresponding loss curves for the DNN model. Model performances are shown in terms of coefficient of determination $R^{2}$, mean absolute error (MAE), and root mean squared error (RMSE).}
    \label{fig:newton_model}
\end{figure}

In this case, our aim is to identify the invariance of the noised $\overline{F_{\mathrm{g}}}$ with respect to groups $(x_1 - x_2)$, $(y_1 - y_2)$, $(z_1 - z_2)$. To this end, we employ the procedure presented in subsection \ref{general} for identification of general group form, focusing specifically on groups with functional dependence $\alpha_1 x_i+\alpha_2 x_j$.
%
As usual, we get access to the gradient $\nabla \overline{F_\mathrm{g}}(x_1, x_2, y_1, y_2, z_1, z_2, m_1, m_2)$ by means of automatic differentiation over a DNN approximating $\overline{F_\mathrm{g}}(x_1, x_2, y_1, y_2, z_1, z_2, m_1, m_2)$.
Fig.~\ref{fig:newton_model}a shows the model predictions over the testing set, while in Fig.~\ref{fig:newton_model}b the corresponding loss across epochs is depicted; specifically, the model is highly predictive, with coefficient of determination $R^2=0.959$ and no evidence of overfitting is observed. 
The trained model is thus fed to the optimization algorithm. The average $\mu_{\alpha_1}, \mu_{\alpha_2}$ estimates of the coefficients $\alpha_1, \alpha_2$, together with their standard deviations $\sigma_{\alpha_1}, \sigma_{\alpha_2}$ are reported in \Cref{table7}. Specifically, the procedure correctly identifies the groups $(x_1, x_2)$, $(y_1, y_2)$, $(z_1, z_2)$ with estimates $\mu_{\alpha_1}, \mu_{\alpha_2}$ of $(-0.729, 0.680)$, $(-0.715, 0.681)$, $(-0.702, 0.697)$ respectively, compared to the true normalized coefficients $(-0.707, 0.707)$ for all cases.
Table \ref{table7} flags as \emph{found} groups with low standard deviations and as \emph{reliable} all the groups in which none of the evaluated exponents approaches 0 or 1. Specifically, other variables couples show high variance over the corresponding average estimates, meaning that no further invariant group is identified.

\begin{table}
    \centering
    \caption{Normalized exponents $\alpha_1, \alpha_2$ for the Newton's law of universal gravitation, together with their average estimates over 20 evaluations $\mu_{\alpha_1}$, $\mu_{\alpha_2}$, and the corresponding standard deviations $\sigma_{\alpha_1}$, $\sigma_{\alpha_2}$. Found groups/sets refer to low standard deviation, reliable groups refer to average far from 1 and 0.}
    \begin{tabular}{crrrrrrcc}
         \toprule
         \textbf{Group} & $\alpha_1$ & $\alpha_2$ & ${\mu_{\alpha_1}}$ & ${\sigma_{\alpha_1}}$ & ${\mu_{\alpha_2}}$ & ${\sigma_{\alpha_2}}$ & \textbf{Found} & \textbf{Reliable} \\
         \midrule
         $(x_1, x_2)$ & 0.707 & -0.707 & -0.729 & 0.047 & 0.680 & 0.057 & yes & yes \\
         $(y_1, y_2)$ & 0.707 & -0.707 & -0.715 & 0.09 & 0.681 & 0.128 & yes & yes \\
         $(z_1, z_2)$ & 0.707 & -0.707 & -0.702 & 0.103 & 0.697 & 0.104 & yes & yes \\
         $(m_1, m_2)$ & - & - & -0.455 & 0.601 & 0.294 & 0.587 & no & - \\
         $(m_1, x_1)$ & - & - & 0.000 & 0.000 & 1.000 & 0.000 & no & - \\
         $(x_1, y_1)$ & - & - & -0.628 & 0.271 & -0.032 & 0.714 & no & -\\
         \bottomrule
    \end{tabular}
    \label{table7}
\end{table}

\section{Conclusions and final remarks}
In this study, we have implemented two innovative methodologies for searching optimal variables to describe physical data, making use of both regression and classification models applied to the data.
Specifically, leveraged on well-suited datasets for machine learning, with the goal of predicting a property of interest.

More precisely, the procedure based on the regression model introduced here enables the identification of invariant groups and/or sets of variable groups with respect to which the property of interest is invariant. 
We have demonstrated its effectiveness on noised data generated by three well known functional relationships: the Dittus-Boelter correlation, the Gnielinski correlation, and Newton's law of universal gravitation.
For the former two cases, the procedure has accurately detected groups/sets in power form, while for the latter case, a generalized algorithm successfully identified groups with a linear form. 
Interestingly, the methodology is potentially applicable to any functional form, as illustrated in subsection \ref{general}. 

The procedure based on classification of the physical property values, allows the determination of an appropriate set of exponents to combine all the primitive variables in power form, thus constructing mixed features optimized for the classification task.
Specifically, we have shown its effectiveness on the Dittus-Boelter and on the Gnielinski correlations. 
Indeed, the methodology effectively enables the separation of classes even with just one mixed feature, whereas a single primitive variable fails to achieve class separation. 
Furthermore, we also provide examples with one and two mixed variables, together with separation in two or three classes.

It is worth stressing that the methodologies presented in this study are blessed by generality and, as such, are not restricted to the selected case studies. 
Therefore, potential applications can be envisioned in various other fields in the future.
%
In particular, we notice that the identification of effective good variables is not only interesting \emph{per se} to possibly gain a deeper insight on a physical system, but can also be practically advantageous when designing experiments. 
Indeed, the methodology to detect groups/sets in regression facilitates efficient group/set-level adjustments rather than individually fine-tuning variables within the groups/sets themselves.

Moreover, the classification procedure can enable the reduction of the numerous original primitive variables to a minimal set of optimized variables concerning a specific physical property of interest. 
Notably, combinations of variables yielding identical mixed feature values exhibit similar performance in terms of the property being classified. 
As a result, it is possible to find alternative combinations of primitive variables without compromising the overall performance of a given system under study. 
From a more practical standpoint, the methodologies described here can help generalizing optimal system conditions, thus helping decreasing the most resource-intensive components while properly re-balancing the others.
As an example, these general methodologies may thus hold the potential to save resources, such as costly reagents (e.g., Bonke \emph{et al.}~\cite{bonke2023multi} for solar fuel production) or expensive materials as in perovskite solar cells optimization \cite{tailor2020recent}. 
An additional advantage associated to the correct identification of a reduced set of ruling variables is their possible use for driving sequential learning or Bayesian optimization processes \cite{huan2013simulation,motoyama2022bayesian}.   

Finally, we believe that an interesting development of the presented methodologies to be pursued in the future, shall be in the direction of a possible handling of systems ruled not only by numerical parameters but also categorical ones.

\section*{Data availability}
Processed datasets and trained models of this study are publicly available in Zenodo at DOI: 10.5281/zenodo.10406490 


\section*{Code availability}
The codes used to obtain the results of this study are publicly available in github at \texttt{https:// \linebreak github.com/giuliobarl/GoodPhysVariables}.


\clearpage

\end{document}


\maketitle

\subsection*{Supplementary Note 1: Liquids properties}

\begin{table}[H]
    \centering
    \caption{Physical properties (i.e., density $\rho$, dynamic viscosity $\mu$, thermal conductivity $\lambda$, specific heat $c_{\mathrm{p}}$, kinematic viscosity $\nu$, thermal diffusivity $\kappa$) at ambient temperature and pressure of the 16 real liquids used to create the datasets for the fluid correlations Dittus-Boelter and Gnielisnki [1].}
    \begin{tabular}{ccccSSS}
         \toprule
          & \multicolumn{1}{c}{$\rho$} & \multicolumn{1}{c}{$\mu$} & \multicolumn{1}{c}{$\lambda$} & \multicolumn{1}{c}{$c_{\mathrm{p}}$} & \multicolumn{1}{c}{$\nu$} & \multicolumn{1}{c}{$\kappa$} \\
         
         \multicolumn{1}{c}{\textbf{Formula}} & \multicolumn{1}{c}{$(\si{\gram\cdot\milli\liter^{-1}})$} & 
         \multicolumn{1}{c}{$(\si{\milli\pascal\cdot\second})$} 
         & \multicolumn{1}{c}{$\left(\si{\watt\meter^{-1}\kelvin^{-1}}\right)$} &
         \multicolumn{1}{c}{$\left(\si{\joule~\kilo\gram^{-1}~\kelvin^{-1}}\right)$} &
         \multicolumn{1}{c}{$\left(\times 10^{-9}\, \si{\meter^2~\second^{-1}}\right)$} &
         \multicolumn{1}{c}{$\left(\times10^{-9}\, \si{\meter^2~\second^{-1}}\right)$}\\
         
         \midrule
         $\rm{H_2O}$ & 0.998 & 0.893 & 0.607 & 4179.85 & 894.79 & 145.47 \\
         $\rm{CCl_4}$  & 1.584 & 0.908 & 0.099 & 849.69 & 573.10 & 73.16 \\
         $\rm{CHCl_3}$ & 1.480 & 0.537 & 0.117 & 956.61 & 362.91 & 82.79 \\
         $\rm{CH_2Br_2}$ & 2.477 & 0.980 & 0.109 & 604.04 & 395.64 & 72.21 \\
         $\rm{CH_4O}$ & 0.787 & 0.549 & 0.200 & 2531.05 & 697.91 & 100.40 \\
         $\rm{CS_2}$ & 1.256 & 0.352 & 0.149 & 1003.43 & 280.37 & 118.34 \\
         $\rm{C_2Cl_4}$ & 1.623 & 0.845 & 0.110 & 864.74 & 520.64 & 78.59 \\
         $\rm{C_2H_3Cl_3}$ & 1.435 & 0.793 & 0.116 & 1188.73 & 552.61 & 74.79 \\
         $\rm{C_2H_3N}$ & 0.776 & 0.344 & 0.188 & 2226.41 & 443.02 & 108.58 \\
         $\rm{C_2H_4O_2}$ & 1.044 & 1.056 & 0.158 & 2053.23 & 1011.57 & 73.71 \\
         $\rm{C_2H_5Cl}$ & 0.889 & 0.258 & 0.119 & 1616.69 & 290.24 & 82.87 \\
         $\rm{C_2H_6O}$ & 0.785 & 1.074 & 0.169 & 2437.68 & 1368.27 & 88.43 \\
         $\rm{C_3H_5ClO}$ & 1.181 & 1.073 & 0.131 & 1422.33 & 908.40 & 77.85 \\
         $\rm{C_3H_6O}$ & 0.784 & 0.306 & 0.161 & 2174.61 & 390.11 & 94.62 \\
         $\rm{C_3H_6O_2}$ & 0.928 & 0.364 & 0.153 & 1915.54 & 392.28 & 86.30 \\
         $\rm{C_6H_6}$ & 0.874 & 0.604 & 0.141 & 1744.98 & 691.39 & 92.56 \\
         \bottomrule
    \end{tabular}
    \label{tab:table1}
\end{table}

\clearpage

\subsection*{Supplementary Note 2: DNN structure}
\begin{table}[H]
    \centering
    \caption{The architecture of all the DNNs used in this study. All DNNs were trained using the optimizer implementing the Adam algorithm, with a learning rate $ = 0.001$; $n$ denotes the number of input features in each dataset ($n=4$ for Dittus-Boelter dataset, $n=5$ for Gnielinski dataset, $n=8$ for Newton dataset).}
    \begin{tabular}{ccc}
         \toprule
         \textbf{Layer type} & \textbf{Dimensions} & \textbf{Activation} \\
         \midrule
         Input & $n$ & - \\
         Normalization & - & - \\
         Dense & 128 & ReLU \\
         Dense & 64 & ReLU \\
         Dense & 64 & ReLU \\
         Dense & 32 & ReLU \\
         Dense & 32 & ReLU \\
         Dense & 16 & ReLU \\
         Dense & 16 & ReLU \\
         Dense & 8 & ReLU \\
         Dense & 1 & linear\\
         \bottomrule
    \end{tabular}
    \label{tab:table2}
\end{table}

\clearpage

\subsection*{Supplementary Note 3: Local invariance}

As outlined in the main text, the regression procedure detects \emph{local} invariances for the Gnielinski correlation. We briefly recap here below the procedure to numerically verify this (groups of two variables only).

Firstly, the features not present in the group are held constant. Secondly, a random sample $i$ from the dataset is chosen to evaluate the quantity $\Tilde{c}=x_{1,i}^{\alpha_1} x_{2,i}^{\alpha_2}$. Thirdly, a vector $\overline{\mathbf{x}_{1}}$ is generated, consisting of evenly spaced points in the domain of the first feature. Consequently, the array corresponding to the second feature is derived as $\overline{\mathbf{x}_2} = (\Tilde{c}\overline{\mathbf{x}_1}^{-\alpha_1})^{1/\alpha_2}$. A new dataset is then formed, where variables outside the group remain constant, and those within the group are replaced by $\overline{\mathbf{x}_1}$ and $\overline{\mathbf{x}_2}$. For all these samples, the response value $f(\mathbf{x})$ is computed. The local invariance is demonstrated when $f(\mathbf{x})$ remains approximately constant across the newly constructed dataset.

\begin{figure}[H]
    \centering
    \includegraphics[width = \textwidth]{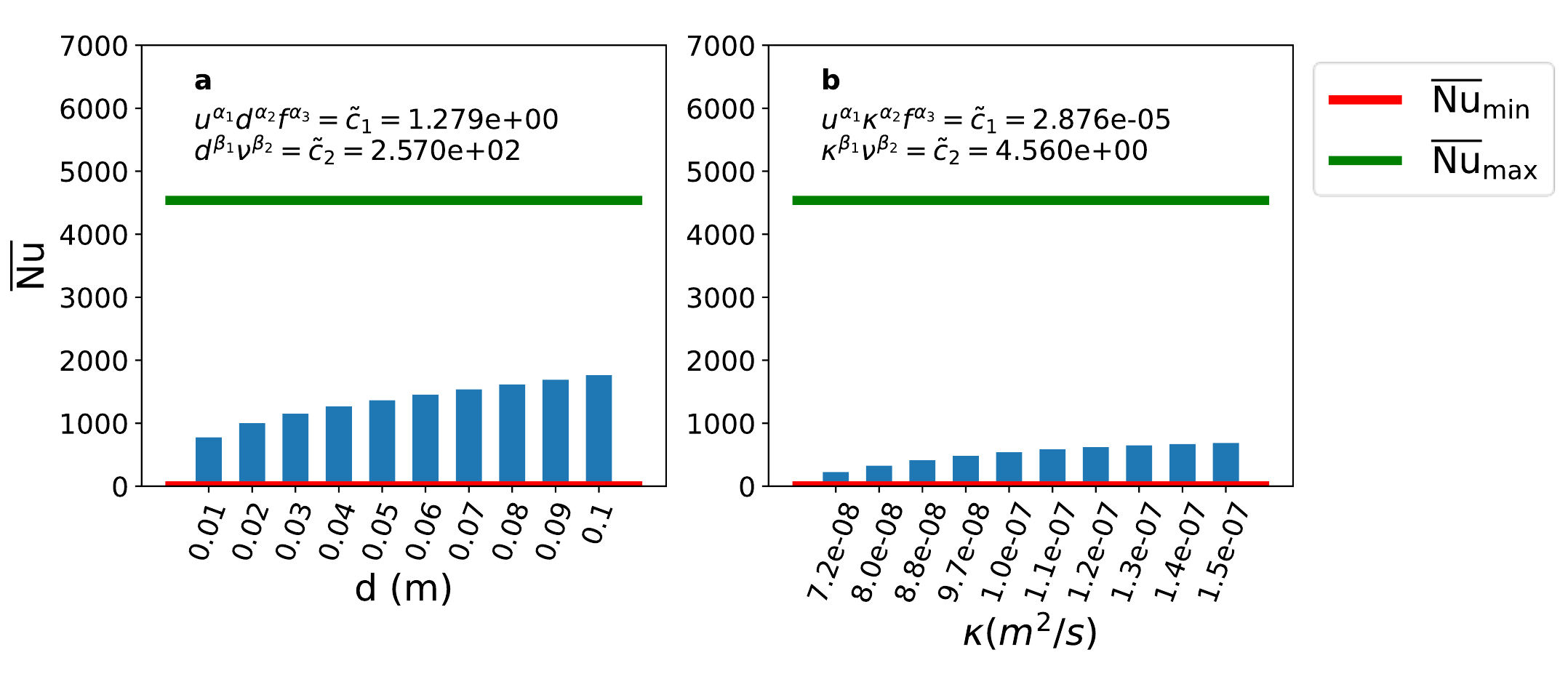}
    \caption{Results of the numerical validation of the \textit{local} invariance for sets of groups in the power form of triplet and couple. \textbf{a} validation for the case $u^{\alpha_1}d^{\alpha_2}f^{\alpha_3}$ and $d^{\beta_1}\nu^{\beta_2}$. \textbf{b} validation for the case $u^{\alpha_1}\kappa^{\alpha_2}f^{\alpha_3}$ and $\kappa^{\beta_1}\nu^{\beta_2}$.}
    \label{fig:local-sets}
\end{figure}

In our case, $f(\mathbf{x})$ corresponds to the noised Nusselt number $\overline{\mathrm{Nu}}$. As depicted in Supplementary Figs.~\ref{fig:local}, \ref{fig:local-couples}, \ref{fig:local-sets}, over the examined groups in the power forms of couples, set of two couples, set of a triple and a couple respectively, the local invariance is always verified. Notably, the maximum and the minimum values $\overline{\mathrm{Nu}}_{\mathrm{min}}$, $\overline{\mathrm{Nu}}_{\mathrm{max}}$ over the whole original dataset 
are appreciably different if compared with the \emph{invariant} $\overline{\mathrm{Nu}}$ value over the new dataset. 

\begin{figure}[H]
    \centering
    \includegraphics[width = \textwidth]{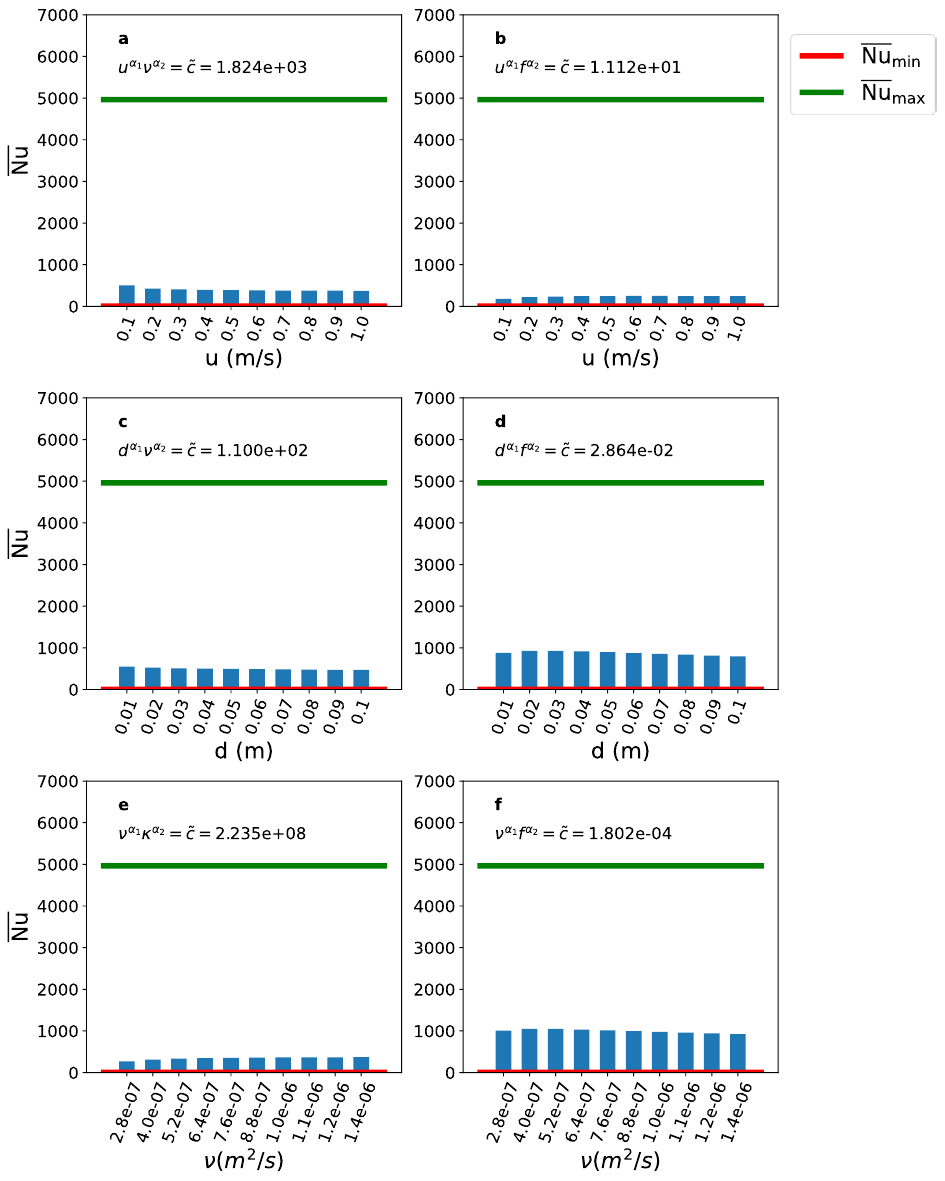}
    \caption{Results of the numerical validation of the \textit{local} invariance for groups in the power form of couple. \textbf{a} validation for the case $u^{\alpha_1}\nu^{\alpha_2}$. \textbf{b} validation for the case $u^{\alpha_1}f^{\alpha_2}$. \textbf{c} validation for the case $d^{\alpha_1}\nu^{\alpha_2}$. \textbf{d} validation for the case $d^{\alpha_1}f^{\alpha_2}$. \textbf{e} validation for the case $\nu^{\alpha_1}\kappa^{\alpha_2}$. \textbf{f} validation for the case $\nu^{\alpha_1}f^{\alpha_2}$.}
    \label{fig:local}
\end{figure}

\begin{figure}[H]
    \centering
    \includegraphics[width = \textwidth]{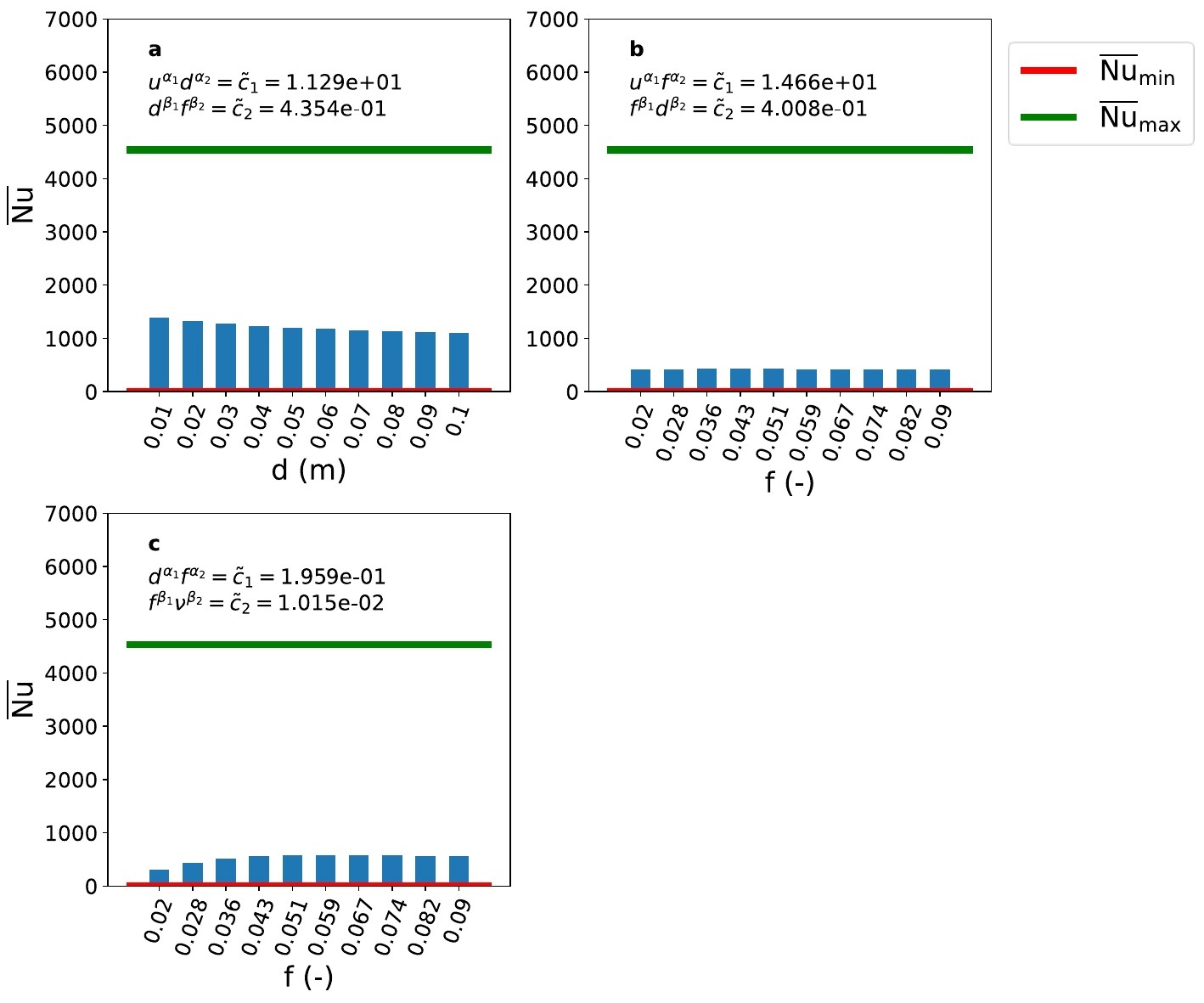}
    \caption{Results of the numerical validation of the \textit{local} invariance for sets of groups in the power form of couple. \textbf{a} validation for the case $u^{\alpha_1}d^{\alpha_2}$ and $d^{\beta_1}f^{\beta_2}$. \textbf{b} validation for the case $u^{\alpha_1}f^{\alpha_2}$ and $f^{\beta_1}d^{\beta_2}$. \textbf{c} validation for the case $d^{\alpha_1}f^{\alpha_2}$ and $f^{\beta_1}\nu^{\beta_2}$.}
    \label{fig:local-couples}
\end{figure}

\clearpage

\subsection*{Supplementary Note 4: Mixed optimized variables for classification}
\begin{table}[H]
    \centering
    \caption{Results for variables optimization in classification. \#mixed variables represents the number of optimized variables constructed by mixing the primitive normalized physical variables in power combination form; \#classes denotes the count of classes into which the dataset has been split. The primitive normalized physical variables are the normalized flow velocity $u_{\textrm{norm}}$, the normalized hydraulic diameter $d_{\textrm{norm}}$, the normalized kinematic viscosity $\nu_{\textrm{norm}}$, the normalized thermal diffusivity $\kappa_{\textrm{norm}}$, the normalized friction factor $f_{\textrm{norm}}$, finding optimized exponents $\alpha_1, \alpha_2, \alpha_3, \alpha_4, \alpha_5$, respectively.}
    \begin{tabular}{cccrrrrr}
         \toprule
         \textbf{Correlation} & \#mixed variables & \#classes & $\alpha_1$ & $\alpha_2$ & $\alpha_3$ & $\alpha_4$ & $\alpha_5$ \\
         \midrule
         Dittus-Boelter & 1 & 2 & 0.642 & 0.684 & -0.322 & -0.127 & - \\
         Dittus-Boelter & 2 & 2 & 0.179 & 0.894 & -0.399 & -0.098 & - \\
          &  &  & 0.876 & 0.383 & -0.250 & -0.151 & - \\
         Dittus-Boelter & 1 & 3 & 0.678 & 0.652 & -0.305 & -0.150 & - \\
         Gnielinski & 1 & 2 & 0.648 & 0.628 & -0.329 & -0.090 & 0.263 \\
         Gnielinski & 2 & 2 & 0.733 & 0.504 & -0.316 & -0.060 & 0.324 \\
          &  &  & -0.038 & -0.922 & 0.318 & 0.075 & -0.203 \\
         Gnielinski & 1 & 3 & 0.638 & 0.614 & -0.347 & -0.059 & 0.304 \\
         \bottomrule
    \end{tabular}
    \label{tab:table3}
\end{table}

\clearpage

\section*{References}
\begin{enumerate}[label={[\arabic*]}]
    \item Lide, D. R. \& Kehiaian, H. V. CRC handbook of thermophysical and thermochemical data (CRC press, 2020).
\end{enumerate}